\documentclass[%
 reprint,
superscriptaddress,
nofootinbib,
 amsmath,amssymb,
 aps,
 pra,
]{revtex4-2}

\usepackage{graphicx}
\usepackage{dcolumn}
\usepackage{bm}
\usepackage{hyperref}
\usepackage{bbold}

\usepackage{braket}
\usepackage{subfigure}
\usepackage[utf8]{inputenc}
\newcommand{\e}{\mathrm{e}}
\renewcommand{\i}{\mathrm{i}}
\newcommand{\eqdef}{\overset{\text{def.}}{=}}

\newcommand{\id}{\hat{\mathbb{1}}}
\newcommand{\tr}{\mathrm{tr}}
\renewcommand{\d}{\mathrm{d}}

\newcommand{\h}{\hat{H}}
\newcommand{\ph}{\hat{h}}
\newcommand{\hint}{\hat{H}_{\text{int}}}
\renewcommand{\a}{\hat{a}}
\newcommand{\ad}{\hat{a}^{\dagger}}
\newcommand{\ak}{\hat{a}_k}
\newcommand{\akd}{\hat{a}_k^{\dagger}}
\newcommand{\bk}{\hat{b}_k}
\newcommand{\bkd}{\hat{b}_k^{\dagger}}

\newcommand{\hc}{\text{h.c.}}
\newcommand{\proj}[1]{\ket{#1}\bra{#1}}

\begin{document}

\title{Unveiling non-Markovian spacetime signalling in open quantum systems with long-range tensor network dynamics}

\author{Thibaut Lacroix}
\email{tfml1@st-andrews.ac.uk}
\affiliation{%
 SUPA, School of Physics and Astronomy, University of St Andrews, St Andrews KY16 9SS, UK}
 \affiliation{
 Sorbonne Universit\'{e}, CNRS, Institut des NanoSciences de Paris, 4 place Jussieu, 75005 Paris, France
}%

\author{Angus Dunnett}
 \affiliation{
 Sorbonne Universit\'{e}, CNRS, Institut des NanoSciences de Paris, 4 place Jussieu, 75005 Paris, France
}%

\author{Dominic Gribben}
\affiliation{%
 SUPA, School of Physics and Astronomy, University of St Andrews, St Andrews KY16 9SS, UK}

\author{Brendon W. Lovett}
\affiliation{%
 SUPA, School of Physics and Astronomy, University of St Andrews, St Andrews KY16 9SS, UK}

\author{Alex Chin}
 \affiliation{
 Sorbonne Universit\'{e}, CNRS, Institut des NanoSciences de Paris, 4 place Jussieu, 75005 Paris, France
}%

\date{\today}

\begin{abstract}
Nanoscale devices - either biological or artificial - operate in a regime where the usual assumptions of a structureless, Markovian, bath do not hold.
Being able to predict and study the dynamics of such systems is crucial and is usually done by tracing out the bath degrees of freedom, which implies losing information about the environment.
To go beyond these approaches we use a numerically exact method relying on a Matrix Product State representation of the quantum state of a system and its environment to keep track of the bath explicitly.
This method is applied to a specific example of interaction that depends on the spatial structure of the system.
The result is that we predict a non-Markovian dynamics where long-range couplings induce correlations into the environment.
The environment dynamics can be naturally extracted from our method and shine a light on long time feedback effects that are responsible for the observed non-Markovian recurrences in the eigen-populations of the system.
\end{abstract}

\maketitle


\section{Introduction}\label{sec:intro}

\begin{figure}
    \centering
    \includegraphics[width=\linewidth]{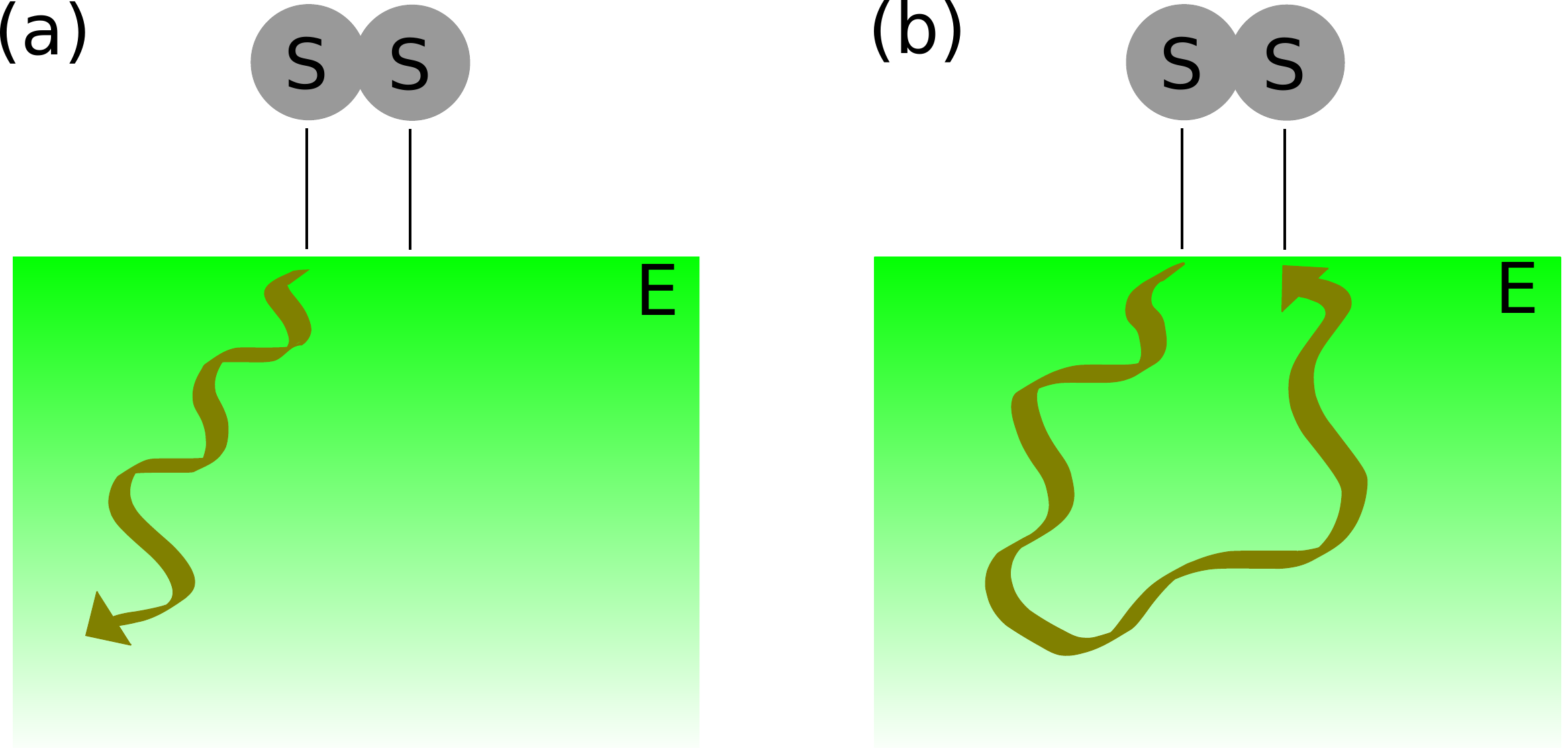}
    \caption{Schematic representation of (a) a Markovian environment and (b) a non-Markovian environment. In a Markovian environment, excitations created through the interaction with the bath propagate away and don't influence the system. By contrast, in the non-Markovian case, these excitations can have a backaction at a later time on a different part of the system.}
    \label{fig:markovian}
\end{figure}

Real life quantum systems are never truly isolated from the rest of the Universe and are typically exposed to a macroscopic number of fluctuating degrees of freedom that constitute their often unobservable - and invariably uncontrollable - environments \cite{Breuer, Weiss}. 
Weak interactions of a quantum system with spectrally broad and dynamically featureless environments lead to so-called \emph{Markovian} dissipation in which energy relaxation and decoherence can be accurately described by a time-local Redfield or Lindblad master equation \cite{Breuer, Weiss, Blum}.
In these `leaky' systems, the perturbations of the environment caused by the system rapidly and irreversibly propagate away, essentially removing any trace, or `memory', of prior interactions in the way the environment acts locally on the embedded system (see Fig.~\ref{fig:markovian}). 
Acting always in the instant and having no dependence on the shared history of the system-bath interactions, Markovian noise is thus very difficult to control, and most strategies to combat its unwanted effects simply aim at its total suppression.

However, in functional nanoscale materials the dividing line between the system and environmental excitations becomes less clear, and large and long-lasting correlations between them can build up over the duration of a process. 
In the presence of these non-equilibrium conditions, these correlations can, \textit{inter alia}, lead to non-classical work extraction, energy transport and violation of detailed balance \cite{Strasberg2016, Giorgi2015, Oviedo2016}. The investigation of how open system-environment correlations influence and might even help optimize energy harvesting, transport and transduction processes in devices operating at the few-quanta level is an important research line in the burgeoning field of quantum thermodynamics \cite{vinjanampathy2016quantum,kosloff2013quantum}.   

\begin{figure}
\includegraphics[width=\columnwidth]{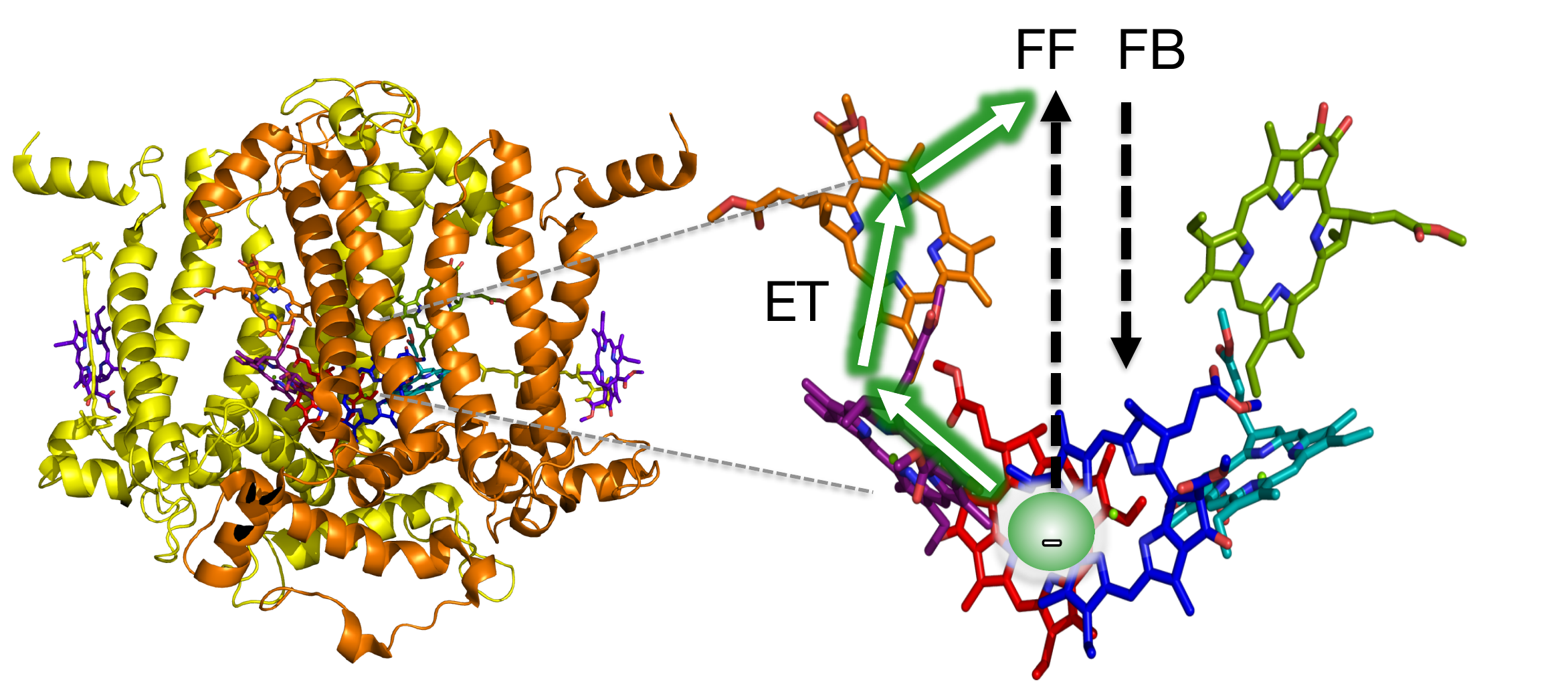}
\caption{Biological inspiration for our correlated bath model. (Left) The protein structure of a nanoscale photosynthetic reaction centre. Photoactive pigments are held rigidly by non-covalent protein interactions that also tune their electronic overlaps, interactions and excited state energies. The coordination of multiple cofactors by extended structures, such as quasi-1d alpha helices, allows vibrational fluctuations to act on different cofactors in a spatio-temporally correlated manner. (Right) Structure of the cofactors active in charge separation through quantum electron transport (ET). The oxidation of water in photosynthesis requires four successful ETs, and this multi-fermion process is regulated through feed-forward (FF) and feedback (FB) mechanisms induced by strong electron-hole interactions with the dissipative protein scaffold.}
\label{fig:allostery}
\end{figure}

Nowhere are these concepts of more relevance than in the protein-based `nanomachines' that Nature has developed to perform the key optoelectronic tasks of photosynthesis. For example, the pigment-protein complexes (PPC) that perform the electron transfers at the core of photosynthesis are composed of photoactive pigments in interaction with a highly structured environment made of a protein scaffold that tunes the electronic and vibrational properties of the molecular network. The structure of such a `reaction center' (RC) is shown in Fig. \ref{fig:allostery}. The electron transport (ET) chain is shown on the RHS of Fig. \ref{fig:allostery}, beginning at the `special pair' of chlorophyll and terminating at the quinone acceptors (not shown). In higher plants, the hole left behind by ET is ultimately refilled by the splitting of water and evolution of oxygen \cite{Blankenship}. This requires the RC is turn over \emph{four} electrons in a concerted action, a remarkable feat of multi-carrier photocatalysis. 

Coordinating multiple charge dynamics in structures with poor dielectric screening and typical lateral sizes of only $5-6$ nm requires exquisite spatio-temporal control of energy transfer and ET, including mechanisms of feedback to ensure the processes occur in the correct order \emph{without} waste of excited state energies. While the role of the structured environments found PPCs has been widely discussed in terms of transport efficiency and the possible support of coherent electronic dynamics in light-harvesting \cite{Engel2007,Collini2010,chin2013role,kreisbeck2012long}, the signalling and potential efficiency gains from spatio-temporal feedback (FB) and heralding feedforward (FF) processes in the environment has received rather scant attention. However, first principles methods based on crystal structures do show that the large secondary protein elements that span the ET chain in the RC could `communicate' the initial and final sites of the ET, and may act to prevent accumulation of further charges \cite{muh2013nonheme}. Elsewhere in biology, the idea of dynamical structural changes as a way to regulate processes is well established, especially in the field of allosteric regulation \cite{bozovic2020real,guo2016protein}.     
Considered as an open quantum system problem, the existence of strong spatio-temporal correlations necessitates a manifestly non-Markovian description of the dynamics, as the key physics is encoded in the retarded `action at a distance' that results from previous system-bath interactions, energy exchange, etc. In this article we develop a model that allows us to explore these effects in a fully quantum mechanical description which opens a route to establishing the phenomenology of non-Markovian dissipation in the regime where system dynamics, relaxation transitions \emph{and} environmental signalling occur on similar timescales. By first identifying and understanding the underlying microscopic physics behind these phenomena, we hope to build up a conceptual base that could be used to \emph{exploit} these effects, including any explicitly non-classical effects, in artificial nanoscale devices.

However, capturing non-Markovian dynamics has proven to be quite challenging because of the large amount of information that usually needs to be kept about the system's dynamics and the large number of (often continuous) modes in the environment which subjects such problems to the \emph{curse of dimensionality}: the number of possible quantum states grows exponentially with the number of modes of the environment.
Moreover, non-Markovian dynamics are also non-perturbative and their study thus requires the use of advanced numerical methods. 
There are two broad approaches to this problem, reduced density matrix methods and wave function approaches.
The former does not keep a microscopic description of the environment.
The only information kept about the environment is its correlation function -- or equivalently its spectral density.
The evolution of the system's density matrix can then be described for example, by an approximate weak coupling master equation \cite{Breuer}, or exactly using a process tensor \cite{Pollock} or a tensor network representation of the influence functional as in the Time Evolving Matrix Product Operator (TEMPO) method \cite{strathearn, gribben_exact_2020}. 
Indeed, a process tensor can be extracted from the TEMPO method~\cite{jorgensen2019exploiting} and this can lead to still more efficient calculations~\cite{fux2021efficient}.
The latter distinct approach relies on a wave-function representation of the isolated joint system and keeps an explicit microscopic description of the environment -- but often with an alternative description of its degrees of freedom.
For example, the Time Evolving Density operator with Orthonormal Polynomials Algorithm (TEDOPA) \cite{Chin2010} maps the continuum of independent modes of the environment into a chain with nearest neighbours couplings.
Alternatively, the Multi-Layer Multi-Configuration Time-Dependent Hartree (ML-MCTDH) method \cite{Meyer2012} relies on a description of the environment degrees of freedom with so called time-dependent single particle functions.
Both the reduced density matrix and wave-function approaches have gained numerical efficiency by using tensor networks ans\"atze as their fundamental objects and exploiting efficient contractions and compression techniques.

In this paper, we present an extension of the TEDOPA method to describe system-bath interactions that are long-ranged even in the mapped chain topology.
These long-ranged interactions come in our model from a spatial dependence of the phases of the coupling coefficients between sites of the system and the environment.
We describe the properties of these new couplings and how they can be integrated in the usual \emph{Matrix Product Operator} (MPO) representation of the Hamiltonian in Sec.~\ref{sec:MPO}.
Notably, with this method the new tensors of the MPO scale with the (small) dimension of the reduced system and are independent of the (large) dimension of the environment.
The time evolution is then performed using a one-site \emph{Time Dependent Variational Principal} (1-TDVP) \cite{Dunnett2021} scheme with a \emph{Matrix Product State} (MPS) representation of the wave-function.
Standard tensor network-based approaches, such as Time Evolving Block Decimation (TEBD) \cite{Vidal2004}, are formulated for local interactions and can treat long-range interactions only at the cost of an increased complexity (by increasing the number of steps needed to perform the time evolution, for example via the use of swap gates for TEBD \cite{Shi2006}), thus increasing its computational cost or decreasing its accuracy.
Putting all these elements together, in Sec.~\ref{sec:results} we demonstrate regimes of the model where long time and even periodic communication between the sites is mediated by the environment.

\section{Methods}

\subsection{Model}
We consider a 1-dimensional chain of $N$ sites $\{\alpha\}$ in a common 1-dimensional bosonic bath with modes characterised by the wave-vectors $k \in [-k_c, +k_c]$, where $k_c$ is the environment cut-off wave-vector.
The environment dispersion relation is given by $\omega_k = |k|c$ with $c$ the speed of the phonons in the bath.
We restrict ourselves to the single excitation subspace of the system described by a Hamiltonian $\h_S$ with nearest neighbour hopping.
\begin{align}
    \h =& \h_S + \h_E + \hint \label{eq:hamiltonian1}\\
    =& \sum_{\alpha = 1}^{N} E_\alpha\proj{\alpha} + \sum_{\alpha = 1}^{N-1}J\left(\ket{\alpha}\bra{\alpha + 1} +\hc\right) \nonumber\\
    &+ \int_{-k_c}^{+k_c}\omega_k\akd\ak \d k + \sum_{\alpha}\proj{\alpha}\int_{-k_c}^{+k_c}(g_k^{\alpha}\ak + \hc)\d k \label{eq:hamiltonian2}
\end{align}
where $\ak$ is the annihilation operator of a bath mode of wave-vector $k$, $g_k^{\alpha} = g_k\e^{\i kr_{\alpha}}$, with $g_k = g_{-k} \in \mathbb{R}$, is the coupling strengths between the system and the bath and $r_\alpha$ is the position of the site $\alpha$.\\

\begin{figure}[h!]
    \centering
    \includegraphics[scale=.3]{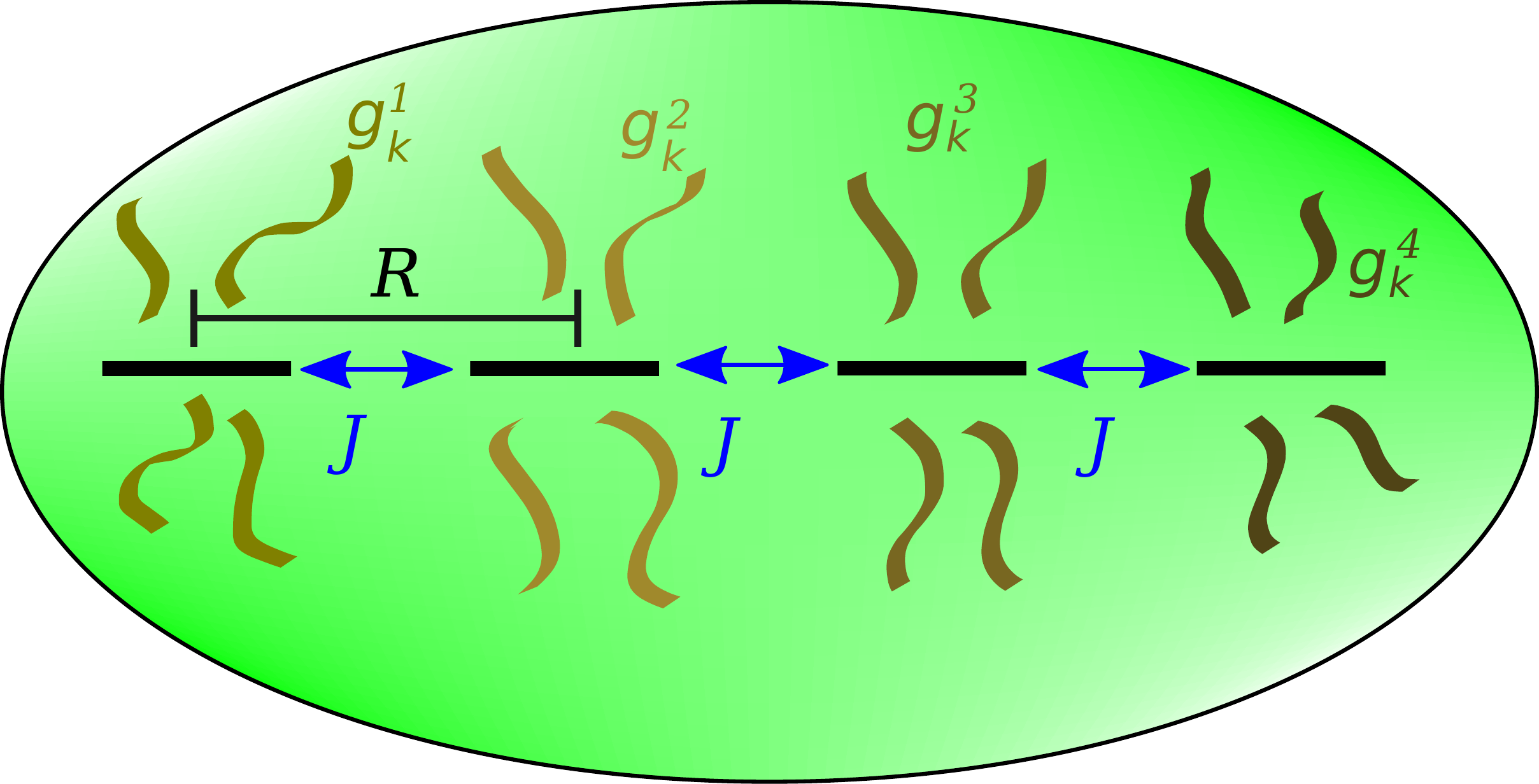}
    \caption{Schematic diagram of the model under study. A system composed of interacting sites is embedded into a single bosonic environment. Each site couples differently to the environment.}
    \label{fig:model_schematics}
\end{figure}

Here, the interaction between the excitation and the bath depends explicitly on the position of this excitation on the chain through the phases of the coupling constants $g_k^\alpha$.
We call this type of coupling a plane-wave coupling.
A schematic of the model is presented in Fig.~\ref{fig:model_schematics}.

In order to study the time evolution of the system and its bath, we use a numerically exact method based on the implementation of the Time Dependent Variational Principal (TDVP) with a tensor network formulation using a MPS ansatz for the quantum states \cite{haegeman}.
This methods requires a discrete representation of the environment in order to write the MPS and to write the Hamiltonian as a MPO.

\subsection{Environment Chain Mapping}
Instead of sampling $k$-modes of the environment to keep only a discrete set of modes, we are using a chain mapping approach that enables us to keep all the relevant bath modes easily and at the same time generate a discrete representation of the environment \cite{Prior2010, Chin2010, Woods2015}.
This method consists of using a unitary transformation defined through a family of orthonormal polynomials that transforms a continuous bosonic environment into a semi-infinite chain and is known as \emph{Time Evolving Density matrix with Orthonormal Polynomials Algorithm} (TEDOPA).

\subsubsection{Zero Temperature}
We separate positive and negative wave-vector modes and apply to them two different chain mappings, and we note $\bk \eqdef \a_{-k}$.
The bath and interaction Hamiltonians become
\begin{align}
        \h_E + \hint = &\int_{0}^{+k_c}\d k \omega_k (\akd\ak +\bkd\bk) \nonumber\\
        &+ \sum_{\alpha}\proj{\alpha}\int_{0}^{+k_c}\d k g_k\left(\e^{\i k r_\alpha}(\ak + \bkd) +\hc \right).
\end{align}

We now introduce two unitary transformations
\begin{align}
     \a_{k\geq 0} &= \sum_n U_n(k) \hat{c}_n\ , \label{eq:UnitaryTransform1}\\
     \hat{b}_{k\geq 0} &= \sum_m V_m(k) \hat{d}_m\ ,
     \label{eq:UnitaryTransform2}
\end{align}
where the matrix elements are
\begin{align}
    U_n(k) = V_n(k) = g_k P_n(k)
\end{align}
where $\{P_n\}_{n \in \mathbb{N}}$ are orthonormal polynomials with respect to the measure $\mu(k) = |g_k^\alpha|^2 = g_k^2 \eqdef J(k)$ (which is the bath spectral density) such that $P_0(k) = 1$ and
\begin{align}
    \int_{0}^{+k_c} P_n(k) P_m(k) J(k)\d k = \delta_{n,m}\ .
    \label{eq:orthogonality}
\end{align}
The nature of the polynomials thus depends on the spectral density of the bath.
They are Jacobi polynomials in the case of an Ohmic spectral density with a hard cut-off (here at $k_c$) $J(k) = 2\alpha k H(k_c - k)$, where $\alpha$ is a coupling strength and $H$ the Heaviside step function.
Another useful property of these polynomials is that they obey a recurrence relation
\begin{align}
    P_n(k) &= (k - A_{n-1})P_{n-1}(k) + B_{n-1}P_{n-2}(k)\ ,
    \label{eq:recurrence}
\end{align}
where $A_n$ is related to the first moment of $P_n$ and $B_n$ to the norms of $P_n$ and $P_{n-1}$ \cite{Chin2010}.
We can then map the bath Hamiltonian using the unitary transformations from Eqs. (\ref{eq:UnitaryTransform1})-(\ref{eq:UnitaryTransform2}) to two tight-binding chains with the same on-site energies $\omega_n$ and hopping energies $t_n$:
\begin{align}
    \h_E &= \sum_{n} \omega_n (\hat{c}^\dagger_n \hat{c}_n + \hat{d}^\dagger_n \hat{d}_n)\nonumber\\
    &+ t_n(\hat{c}^\dagger_n \hat{c}_{n+1} + \hat{c}^\dagger_{n+1} \hat{c}_n + \hat{d}^\dagger_n \hat{d}_{n+1} + \hat{d}^\dagger_{n+1} \hat{d}_n)\ .
\end{align}

For the interaction Hamiltonian, we apply the same procedure and make use of Eq.~(\ref{eq:recurrence}) and find that the chains couple to the system with coupling coefficients $\gamma_n(r_\alpha)$ and $\gamma_n(r_\alpha)^*$
\begin{align}
    \hint &= \sum_{\alpha}\ket{\alpha}\bra{\alpha}\sum_n \Big(\gamma_n(r_\alpha)(\hat{c}_n + \hat{d}^\dagger_{n}) +\hc \Big)
\end{align}
where
\begin{align}
    \gamma_n(r_\alpha) = \int_{0}^{+k_c}\d k J(k)\e^{\i k r_\alpha}P_n(k)\ .
\end{align}

\begin{figure}
\centering
    \subfigure[]{\label{fig:chainmapping}\includegraphics[width=\linewidth]{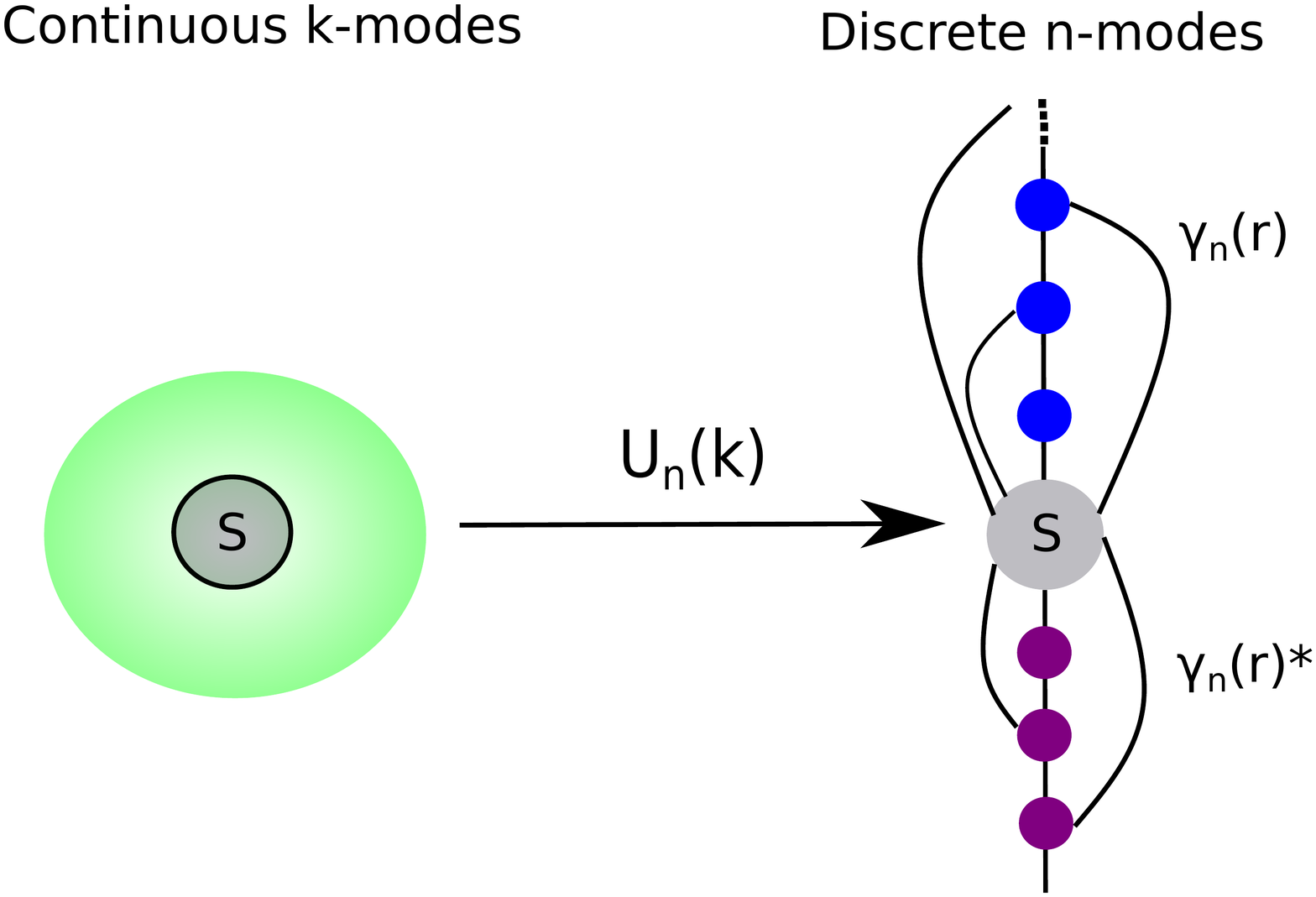}}\\

    \subfigure[]{\label{fig:MPS}\includegraphics[width=\linewidth]{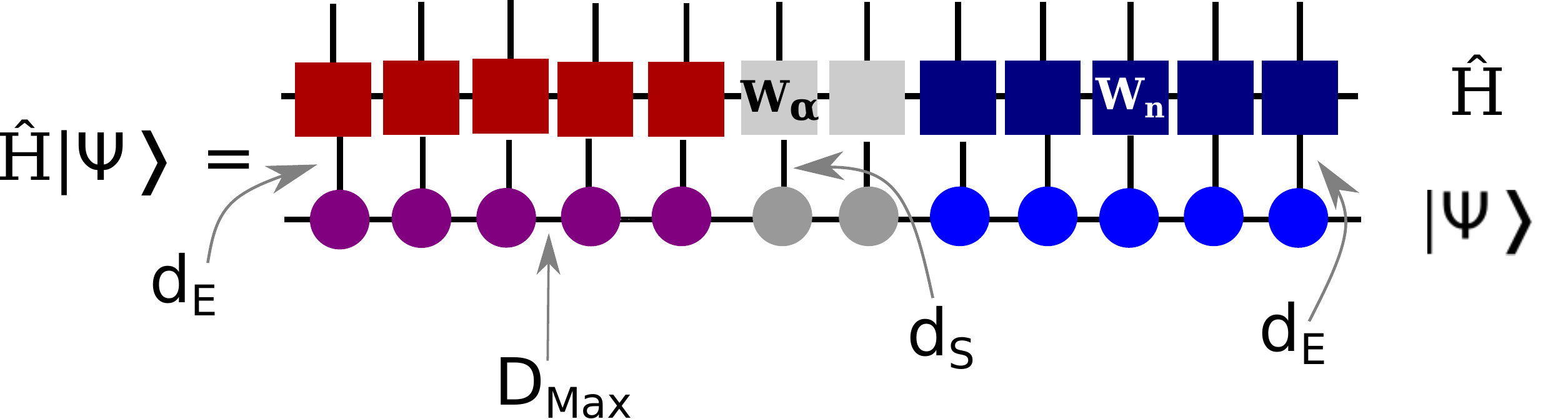}}
\caption{(a) The unitary transformation $U_n(k)$ transforms a continuous environment of uncoupled $k$-modes to semi-infinite discrete tight-binding chains.\\
(b) Schematic diagram of the MPS representation of the wave function of the system and the chain. The circles represent individual tensors which rank is given by their number of \textit{legs}. The open legs correspond to physical Hilbert spaces of dimensions $d_S$ for the system and $d_E$ for the environment. The horizontal legs are virtual bonds related to the amount of correlation between sites, their maximal dimension is $D_\text{Max}$. When a leg is shared between two tensors they are contracted - \textit{i.e.} summed over the corresponding index.}
\end{figure}

In preceding works, TEDOPA resulted in the system being connected only to the first site of the chain.
By contrast, here the system is generally coupled to all the sites of the chain, as represented in Fig.~\ref{fig:chainmapping}.

\subsubsection{Finite Temperature}\label{sec:chain_mapping_finiteT}

This chain mapping technique has been extended to describe finite temperature systems in a statistical mixture as an equivalent zero temperature state vectors under the name \emph{Thermalized - Time Evolving Density matrix with Orthonormal Polynomials Algorithm} (T-TEDOPA) \cite{Tamascelli2019, Dunnett2021}.
It relies on allowing the bath to have negative frequency modes to describe thermal fluctuations and using an alternative bath spectral density that captures the temperature dependence.
To identify this new effective spectral density, we put the finite temperature bath auto-correlation functions $C_\beta(r,t)$ for propagating and contra-propagating modes in the form of a zero temperature auto-correlation $C_\infty(r,t)$.

The interaction Hamiltonian in interaction picture is
\begin{align}
    \h_\text{int}^I &= \sum_{\alpha}\ket{\alpha}\bra{\alpha}\int_{0}^{+k_c}\d k g_k\left(\e^{\i (k r_\alpha - \omega_{k} t)}\a_k +\hc \right) \nonumber\\
    &+\sum_{\alpha}\ket{\alpha}\bra{\alpha}\int_{0}^{+k_c}\d k g_k\left(\e^{-\i (k r_\alpha + \omega_{k} t)}\hat{b}_{k} +\hc \right)\\
    &= \sum_{\alpha}\ket{\alpha}\bra{\alpha}\left(\hat{B}^{1}_{r_\alpha}(t) + \hat{B}^{2}_{r_\alpha}(t) \right)\ .
\end{align}

Hence the bath correlation function for the propagating modes is
\begin{align}
    C_\beta(r-r', t) &= \langle \hat{B}^{1}_r(t) \hat{B}^{1}_{r'}(0) \rangle_B \\
    & = \int_{0}^{+k_c} \d k J(\omega_k) \Big ( n_{\beta}(\omega_k)\e^{-\i(k(r-r') - \omega_k t)} \nonumber\\
    &\ \ \ \ \ \ \ \ \ \ \ \ \ + (n_\beta(\omega_k) +1 )\e^{\i(k(r-r') - \omega_k t)}\Big)
\end{align}
where $n_\beta(\omega_k)$ is the Bose-Einstein distribution and $\beta = (k_B T)^{-1}$ is the inverse temperature.

For zero-temperature, the correlation function reduces to
\begin{align}
    C_\infty(r-r', t) &= \int_{0}^{+k_c} \d k J(\omega_k) \e^{\i(k(r-r') - \omega_k t)} \ .
    \label{eq:CzeroT}
\end{align}

We want to rewrite $C_\beta(r-r', t)$ in the same form as Eq.~(\ref{eq:CzeroT}).
In other words, we want to find a bath at zero $T$ with a different spectral density but with the same system dynamics as the finite $T$ bath.

We recast the first term of $C_\beta(r-r', t)$ such that the argument of the exponential is the same as the second term by sending $k \to -k$ and allowing for negative frequencies.
Hence, $\omega_{-k} = - \omega_k$.
With this transformation we have, in a sense, double the number of propagating modes.
There are the propagating positive $k$ modes with positive energies and the propagating negative $k$ modes with negative energies (coming from the second term of the correlation function).

Finally, the bath correlation function for propagating modes can be written
\begin{align}
     C_\beta(r-r', t) &= \int_{-k_c}^{+k_c} \d k~J_\text{ext}(\omega_{k}) (n_{\beta}(\omega_{k})+1)\e^{\i(k(r-r') - \omega_{k} t)}
\end{align}
with $J_\text{ext}$ is the spectral density with a domain extended to negative frequencies.
The same procedure can be applied to the contra-propagating modes.
We can thus define orthonormal polynomials with the finite-temperature spectral density 
\begin{align}
J_\beta (k) = J_\text{ext}(\omega_{k}) (n_{\beta}(\omega_{k})+1) ,
\end{align}
which is always positive and continuously differentiable.
We define the unitary transformation to chain modes 
\begin{align}
    \a_k &= \sum_n U_n^\beta(k) \hat{c}_n \ \ \text{for} \ \ k \in [-k_c, +k_c]\ ,\\
    \hat{b}_k &= \sum_n U_n^\beta(k) \hat{d}_n \ \ \text{for} \ \ k \in [-k_c, +k_c]
\end{align}
where $U_n^\beta(k) =  \sqrt{J_\beta(k)}P_n^\beta(k)$ and $P_n^\beta(k)$ is a polynomial of order $n$ from a family of orthonormal polynomials with respect to the measure $\d\mu(k) = J_\beta(k)\d k$, i.e. 
\begin{align}
    \int_{-k_c}^{+k_c} P_n^\beta(k) P_m^\beta(k) \d\mu(k) = \delta_{n,m}\ .
\end{align}

With this set of orthogonal polynomials, we can map the environment to two tight binding chains and a coupling coefficient 
\begin{align}
    \gamma_n(r) &= \int_{-k_c}^{+k_c} \d k~J_\beta(\omega_k)\e^{\i k r} P_n^{\beta}(k)\ 
\end{align}
between the system and the $\a_k$ and $\hat{b}^\dagger_k$ operators.

\subsection{Hamiltonian MPO Formulation}
\label{sec:MPO}
To construct the MPO representation of a Hamiltonian $\h$ which is made of a sum of local terms, we use a method based on the recurrence relation presented in \cite{paeckel}.

To define the $k$\textsuperscript{th} tensor of the MPO, we have to decompose the Hamiltonian  into a part that describes what happens before the bond $k$ (which is the bond connecting site $k$ and site $k+1$) $\h_{k-1}^{L}$, after the bond $k$ $\h_{k+1}^R$ and at bond $k$ $\sum_a \ph^L_{k\ a}\otimes\ph^R_{k\ a}$
\begin{equation}
    \label{eq:partition}
    \h = \h_{k-1}^{L}\otimes\id^R_k + \id_k^L\otimes\h_{k+1}^R + \sum_a \ph^L_{k\ a}\otimes\ph^R_{k\ a}
\end{equation}
where $\id^R_k = \underbrace{\id\otimes\ldots\otimes\id}_{N-k+1\text{~times}}$ and $\id^L_k = \underbrace{\id\otimes\ldots\otimes\id}_{k\text{~times}}$.
The last term of Eq.~(\ref{eq:partition}) is an interaction Hamiltonian between the part of the system on the left of bond $k$ and the one on the right of bond $k$.
Hence $\ph^L_{k\ a}$ contains an operator defined on the left of $k$ and $\ph^R_{k\ a}$ an operator defined on the right of $k$ (\textit{e.g.} for a $XYZ$-Hamiltonian with nearest neighbours couplings, we could have $\ph_{k\ a}^L = J_a\hat{S}^a_{k}$ and $\ph_{k\ a}^R = \hat{S}^a_{k+1}$ with $a \in \{x, y,z\}$).
A recurrence relation between the right parts of the Hamiltonian at two consecutive sites can be defined:

\begin{equation}
    \label{eq:recurrenceMPO}
    \begin{pmatrix}
    \h^R_k \\ \ph^R_k\\ \id^R_k
    \end{pmatrix} =
    W_{k+1}
    \begin{pmatrix}
    \h^R_{k+1} \\ \ph^R_{k+1}\\ \id^R_{k+1}
    \end{pmatrix}\ ,
\end{equation}
with the matrices $W_k$ defining the Hamiltonian MPO 
\begin{align}
    \h &= \sum_{\{\sigma\}, \{\sigma^{'}\}, \{w\}} W^{\sigma_1\sigma^{'}_1}_{1\ w_1}W^{\sigma_2\sigma^{'}_2}_{2\ w_1 w_2}\ldots  W^{\sigma_N\sigma^{'}_N}_{N\ w_{N-1}} \ket{\sigma_1\ldots\sigma_N}\bra{\sigma_1'\ldots\sigma_N'}\ . 
    \label{eq:MPO}
\end{align}
In Eq.~(\ref{eq:MPO}) the $\sigma$ and $\sigma'$ indices refer to the local Hilbert spaces of the different parts of the system (\textit{i.e.} sites and chains modes) whereas the $w$ indices relate to virtual bonds between the different parts of the system.
The bath modes will be considered as extra sites where different kind of excitations (which couple to the excitation living on the sites with the $\gamma_n^{\alpha} \eqdef \gamma_n(r_\alpha)$ coefficients) can live.
We introduce a new set of commuting operators $\{\hat{f}_\alpha\}$ such that $\ket{\alpha+1}\bra{\alpha} = \hat{f}_{\alpha+1}^\dagger\hat{f}_{\alpha}$.
Figure \ref{fig:MPS} shows a schematic diagram of the MPO and how it contracts with a MPS.
The on-site tensor has a bond dimension $D = 2(\alpha + 2)$ for the $\alpha$\textsuperscript{th} site and a physical dimension (dimension of the local Hilbert space) $d_S = 2$.
  \begin{equation}
    W_{1} = \left(\id\ \ J_{12}\hat{f}_1\ \ J_{12}\hat{f}_1^\dagger\ \ \ket{1}\bra{1}\ \ \ket{1}\bra{1}\ \ E_1 \proj{1}\right)
  \end{equation}
and
  \begin{align}
   &W_{1 < \alpha \leq N} =\nonumber\\
    &\scalebox{0.9}{$\begin{pmatrix}
    \id & J_{\alpha+1\alpha}\hat{f}_{\alpha} & J_{\alpha+1\alpha}\hat{f}_{\alpha}^\dagger &  0 & 0 & \overbrace{\ldots}^{2(\alpha -2)} & \ket{\alpha}\bra{\alpha} & \ket{\alpha}\bra{\alpha} & E_{\alpha}\proj{\alpha}\\
     &  &  & 0 &  &  &  & & \hat{f}_{\alpha}^\dagger\\
     &  &  & 0 &  &  &  & & \hat{f}_{\alpha}\\
     &  &  & \id &  &  &  & & 0\\
     &  &  &  & \id &  &  & & 0\\
     &  &  &  &  & \ddots & & & \vdots\\
     &  &  &  &  &  & 0 & 0 & 0\\
     &  &  &  &  &  &  & & \id
    \end{pmatrix}$}
\end{align}
with $J_{N+1\ N} = 0$ for the last system tensor.
The chain on-site tensor has a similar structure, but with a constant bond dimension for each mode.
The on-site tensor has a bond dimension $D = 2(N + 2)$ and, in principle, a physical dimension $d = \infty$ that we truncate to a value $d_E$ in our numerical treatment.
The number of sites of the two semi-infinite chains are also truncated at large enough values $N_m$ and $N_m'$, such that an excitation on the chain does not have the possibility to reach the end of the chain during the time evolution
  \begin{equation}
   W_{1\leq n \leq N_m} =
    \begin{pmatrix}
    \id & t_{n}\hat{c}_{n}^\dagger & t_{n}\hat{c}_{n} & 0 & 0 & \ldots & 0 & \omega_{n}\hat{c}_{n}^\dagger\hat{c}_{n}\\
     &  &  &  0 &  &  &  & \hat{c}_{n}\\
     &  &  &  0 &  &  &  & \hat{c}_{n}^\dagger\\
     &  &  & \id &  &  &  & \gamma_{n}^{1}\hat{c}_{n}\\
     &  &  &  & \id &  &  & \gamma_{n}^{1*}\hat{c}_{n}^\dagger\\
     &  &  &  &  & \ddots &  & \vdots\\
     &  &  &  &  &  & \id & \gamma_{n}^{N*}\hat{c}_{n}^\dagger    \\
     &  &  &  &  &  &  & \id
    \end{pmatrix}
    \ ,
\end{equation}
with $t_{N_m} = 0$.
The second chain tensors are identical with $\hat{d}_{n'}$ and $\gamma_{n'}(r)\hat{d}_{n'}^\dagger$ instead of $\hat{c}_n$ and $\gamma_n(r)\hat{c}_n$, where $n'$ corresponds to `mirror' site on the other chain.
The last tensor is
\begin{equation}
    W_{N_m'} =
    \begin{pmatrix}
    \omega_{N_m'}\hat{d}_{N_m'}^\dagger\hat{d}_{N_m'} \\
    \hat{d}_{N_m'} \\
    \hat{d}_{N_m'}^\dagger\\
    \gamma_{N_m'}^{*}\hat{d}_{n'}^\dagger\\
    \gamma_{N_m'}^{1*}\hat{d}_{n'}\\
    \vdots\\
    \gamma_{N_m'}^{N*}\hat{d}_{n'}    \\
    \id
    \end{pmatrix}
    \ .
\end{equation}

One might notice that the chain sites tensors have a bond dimension $D$ that is fixed by the number of sites in the system $N$.
This means that having a large environment only increases the number of individual tensors one needs but not their size.
This result is central for the tractability of this approach.
The identity operators present on the diagonals carry out along the chain the long range coupling coefficients such that they are associated with the corresponding system site.
Hence, they allow a local representation of the Hamiltonian as a MPO even though the interactions are long range across the chain.

To illustrate how the Hamiltonian is recovered from these tensors, we perform the calculation in the case where there is only one site in the system and two modes on a unique chain.
In that case there are only three tensors:
\begin{align}
    W_1 &= 
    \begin{pmatrix}
    \id & \proj{1} & \proj{1} & E_1\proj{1}
    \end{pmatrix}\ ,\\
    W_2 &= 
    \begin{pmatrix}
    \id & t_1\hat{c}_1^\dagger & t_1\hat{c}_1 & 0 & 0 & \omega_1 t_1\hat{c}_1^\dagger\hat{c}_1\\
    0 & 0 & 0 & \id & 0 & \gamma_1^1\hat{c}_1\\
    0 & 0 & 0 & 0 & \id & \gamma_1^{1*}\hat{c}_1^\dagger\\
    0 & 0 & 0 & 0 & 0 & \id
    \end{pmatrix}\ , \\
    W_3 &=
    \begin{pmatrix}
    \omega_2\hat{c}_2^\dagger\hat{c}_2\\
    \hat{c}_2\\
    \hat{c}_2^\dagger\\
    \gamma_2^1\hat{c}_2\\
    \gamma_2^{1*}\hat{c}_2^\dagger\\
    \id
    \end{pmatrix}\ .
\end{align}

The contraction of $W_2$ and $W_3$ gives a $5\times1$ tensor - the same shape as $W_3$ and the transpose of the shape of $W_1$
\begin{align}
    W_2 \cdot W_3 & = 
    \begin{pmatrix}
    \omega_2\hat{c}_2^\dagger\hat{c}_2 + t_1(\hat{c}_1^\dagger\hat{c}_2 + \hat{c}_1\hat{c}_2^\dagger) + \omega_1\hat{c}_1^\dagger\hat{c}_1\\
    \gamma_2^1\hat{c}_2 + \gamma_1^1\hat{c}_1\\
    \gamma_2^{1*}\hat{c}_2^\dagger + \gamma_1^{1*}\hat{c}_{1}^\dagger\\
    \id
    \end{pmatrix}\ .
\end{align}

Further contraction with $W_1$ gives a `scalar' corresponding to the Hamiltonian
\begin{align}
    W_1\cdot W_2 \cdot W_3  =& ~\omega_2\hat{c}_2^\dagger\hat{c}_2 + t_1(\hat{c}_1^\dagger\hat{c}_2 + \hat{c}_1\hat{c}_2^\dagger)+ \omega_1\hat{c}_1^\dagger\hat{c}_1  \nonumber\\
    & + \gamma_2^1\proj{1}\hat{c}_2 + \gamma_1^1\proj{1}\hat{c}_1 + \gamma_2^1\proj{1}\hat{c}_2  \nonumber\\
    & + \gamma_1^{1*}\proj{1}\hat{c}_{1}^\dagger +  \gamma_2^{1*}\proj{1}\hat{c}_2^\dagger\nonumber\\
    &+ E_1\proj{1}\\
    W_1\cdot W_2 \cdot W_3 =& ~\h \ .
\end{align}

In the following, we consider a system made of two degenerate sites with an initial state where the system and its environment are decoupled and the bath is empty
\begin{align}
    \ket{\psi(t=0)} &= \ket{S(0)}\bigotimes_{k\in [-k_c,\ k_c]}\ket{0_k} = \ket{S(0)}\bigotimes_{n \in \mathbb{N}}\ket{0_n}\ ,
\end{align}
where $\ket{S(0)}$ is the initial state of the system and $\ket{0_{k}}$ (respectively $\ket{0_{n}}$) represents the vacuum state of the mode $k$ ($n$) of the bath (the chain).

Adding extra system sites does not add any complexity, but for the sake of clarity in this paper we only present results for two sites.

\section{Zero Temperature}
\label{sec:results}
\subsection{Couplings}
\label{sec:couplings_T=0}
\begin{figure}[h!]
    \centering
    \includegraphics[scale=.4]{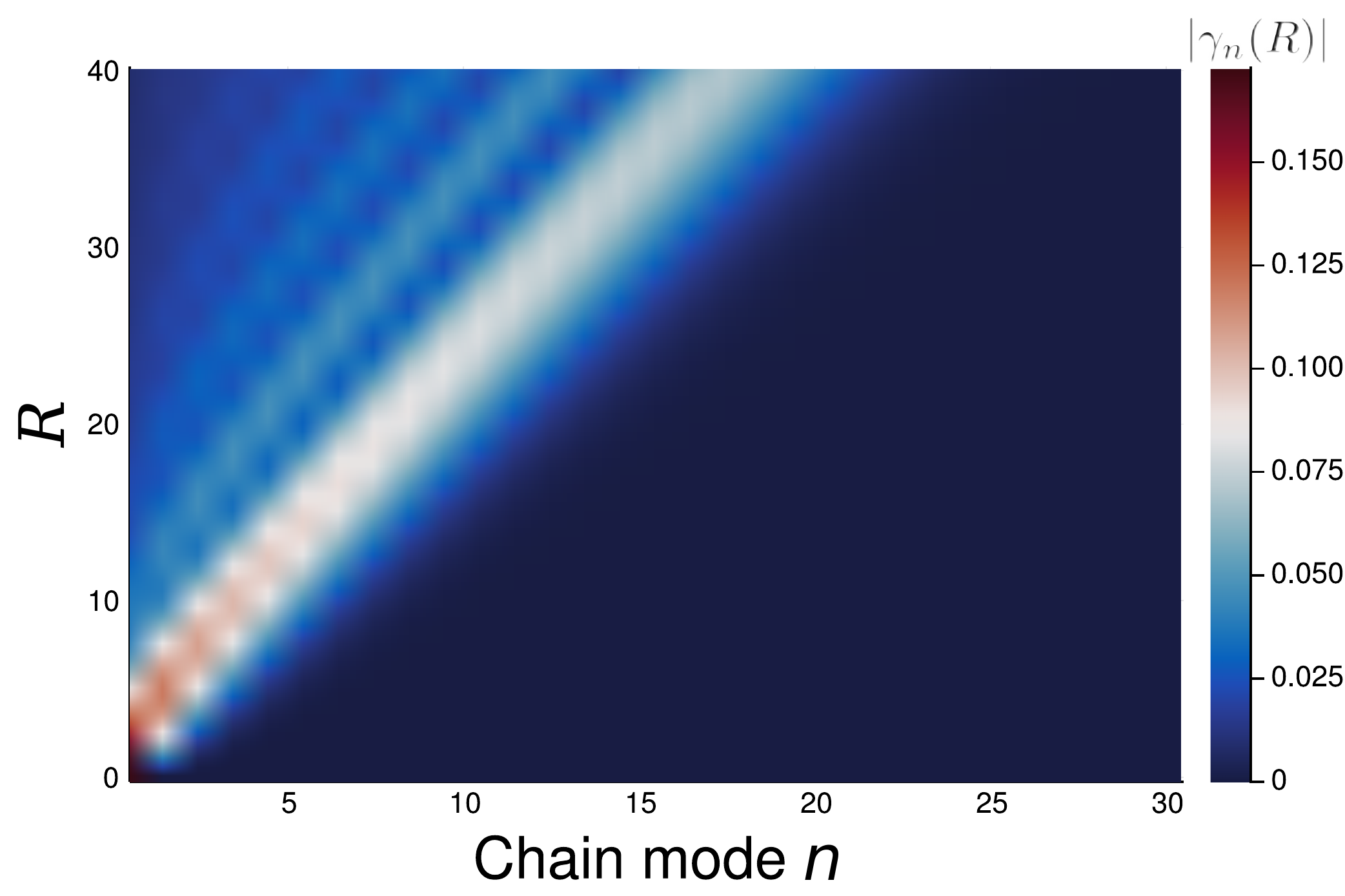}
    \caption{Absolute value of the system-chain coupling constants, for a bosonic bath with a hard cut-off Ohmic spectral density, as a function of the chain modes $n$ and the sites separations $R$. Note that the main peak is centered around $R/2c$. Here $\alpha = 0.12$, $c=1$ and $k_c = 1$.}
    \label{fig:abs_couplings}
\end{figure}

Because of the dependence of the system-bath coupling strengths on the spatial configuration of the system, the system-chain couplings are long-ranged.
In the cases presented in previous works \cite{Tamascelli2019, Dunnett2021} the system only coupled to the first site of the semi-infinite chain.
The system could thus only inject excitation at one end of the chain which then would propagate according only to the tight-binding interactions along the chain.
In the present case, the system-chain couplings are long range and thus the system can create excitations on different regions of the chain.
Absolute values of the system-chain coupling for zero temperature are shown in Fig.~\ref{fig:abs_couplings} for an Ohmic spectral density.

The first site of the system couples only to the first site of the chain.
However the other sites couple to a range of modes with a maximum strength for the mode $n \sim R/2c$ with $R$ the distance between the considered system's site and the first system's site in units of $k_c^{-1}$.  

We can also see in Fig.~\ref{fig:abs_couplings} that the amplitude of the coupling before the peak decreases with the position of the peak.
Said differently, the larger distance between the two sites, the less the second site interacts with the beginning of the chain.
Thus, we can expect that for infinite separation when $R\to\infty$ this system will behave like a Spin-Boson Model (SBM).
This limit is looked at in Appendix \ref{sec:convergence}.

Looking at the opposite limit, when the separation between the two system's site vanishes, Eq.~(\ref{eq:hamiltonian2}) tells us that the system completely decouples from the environment.
Because each site in the system couples mostly to a specific region of the chain, we call our model \emph{``Correlated Environment''} in contrast with the cases where the system couples only to the first site of the chain.

\subsection{Non-Markovian recurrences and bath feedback}
\label{sec:System_T=0}
\begin{figure}[h!]
    \centering
    \includegraphics[width=\columnwidth]{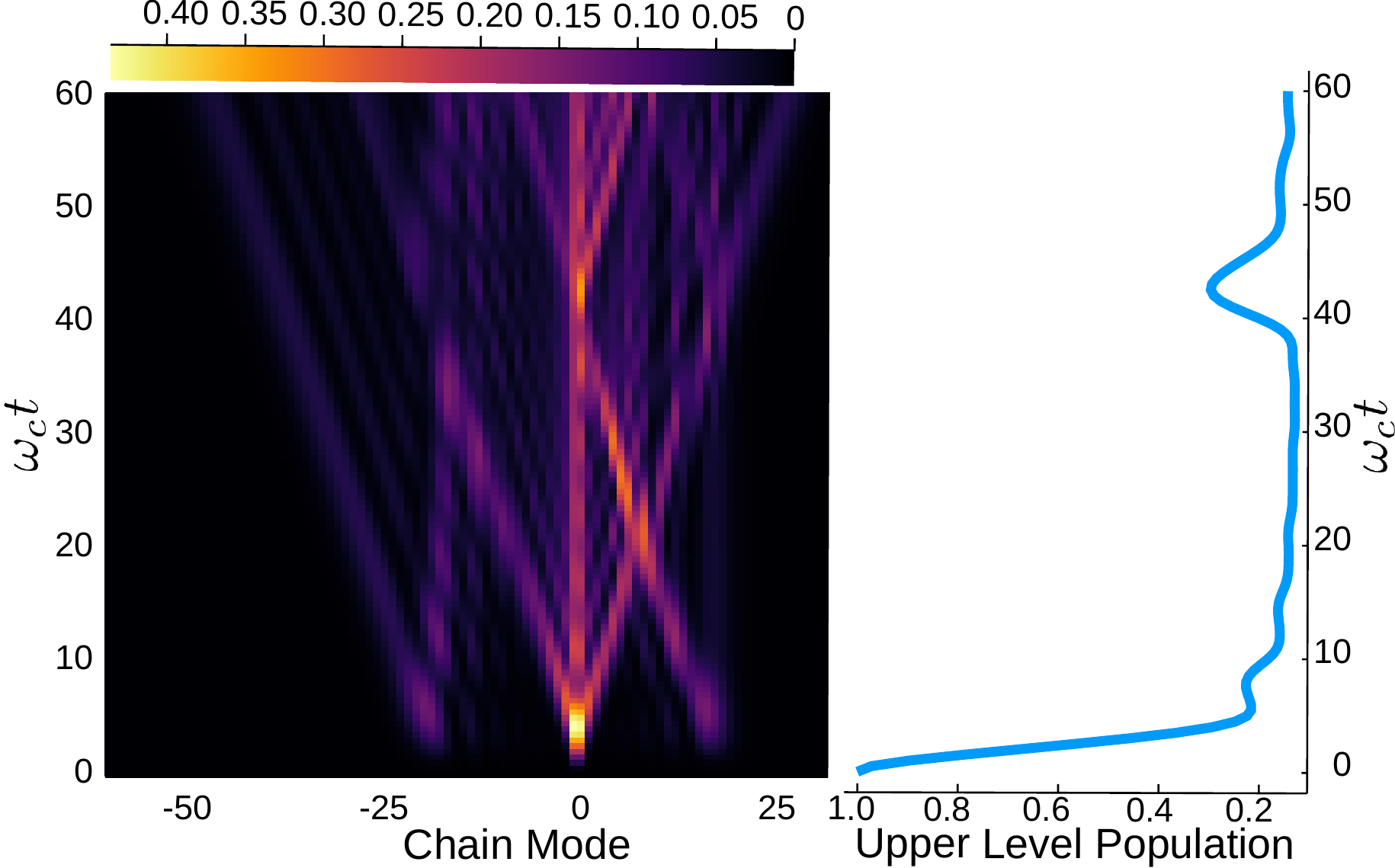}
    \caption{System and bath dynamics. (Left) A heatmap of the chain occupation in time showing the propagation of bath excitations along the chains. (Right) Upper eigenstate population. An eigenstate revival and a site localisation are associated with a chain excitation reaching the beginning of the chain. The separation between the two sites is $R = 40$, their coupling is $J = 0.25$, the speed of sound is $c = 1$, $\alpha = 0.12$ and $k_c = 1$.}
    \label{fig:Upper_R=40_c=1_T=0}
\end{figure}

At zero temperature, the dynamics of the TLS in a bosonic environment is well known and described by the SBM \cite{Breuer}.
In the system's eigen-basis, the population of the upper state (high energy state) should spontaneously decay to the lower state on a time-scale given by the intensity of the coupling between the system and the bath.
The right panel of Figure \ref{fig:Upper_R=40_c=1_T=0} shows the evolution of the eigen-populations with an initial state of the system being the upper eigenstate.
We clearly see that the upper level population decays as expected until $\omega_c t \approx R/c$ when a revival happens. 
This revival corresponds to an increased localisation of the excitation on the second site of the system after following an evolution in a spatial superposition.
With the same conditions, the SBM exhibits the same dynamics except for the revival.
However, we note that the two sites case presented here can be mapped to a SBM with an effective spectral density depending on $R$ (see Appendix \ref{sec:SBMmapping}) but this property is `accidental' and does not generalise to larger systems.

The study of the bath in the chain representation allows us to have a spatial interpretation of the interaction between the system and its environment as the maximum coupling between a system's site and the chain is localized around $n = R/2c$.
The left part of Fig.~\ref{fig:Upper_R=40_c=1_T=0} shows a heatmap of the occupation of the modes of the chains as a function of time.
The positive and negative chain modes each correspond to one of the two chains necessary to take into account propagating and contra-propagating $k$-modes.
The corresponding initial system state is an excitation delocalised on the two sites with a separation $R = 40$.

We can see that the chain modes around $n = \pm R/2c = \pm 20$ get populated first and that the corresponding bath's excitations then propagate on the chains. 
At $\omega_c t \approx 20$ an excitation propagating from the mode $n=0$ coupled mostly to the the first site and an excitation propagating from the mode $n=20$ constructively interfere around $n = 10$.
The former continues to propagate on the chain and traces a ray in the diagram.
The latter reaches $n = 0$ at $\omega_c t \approx 40$ and is reflected.
We can see from this diagram that revivals happen when the excitation emitted along the chain by one site reaches the part of the chain interacting with the other site.
We thus have a feedback effect of the environment on the system.

The dynamics of the chain with negative modes is not the reflection of the dynamics of the chain with positive modes.
Indeed the negative chain modes correspond to the propagating $k$-modes, hence the excitations created by the second site move away from the origin of the chain (which is coupled to the first site).
On the contrary, bath's excitations created by the second site on the positive modes chain correspond to the contra-propagating $k$-modes and move toward the origin of the chain.
On both chains the excitations created by the first system site propagate toward the end of the chain as they move away in real space from the first site.
This explains the apparent `asymmetry' between the two chains.

The dynamics of the system, all other parameters being the same, only depends on the ration $R/c$.
This is also true for the chain dynamics, for example the $(R=40, c=1)$ and $(R=20, c=0.5)$ cases have the same time-frequency diagrams.
This was expected as the system's sites couple in both cases to the same parts of the chain and the bath's excitations travel on the chain at the same speed.

\begin{figure}[h!]
    \centering
    \includegraphics[width=\columnwidth]{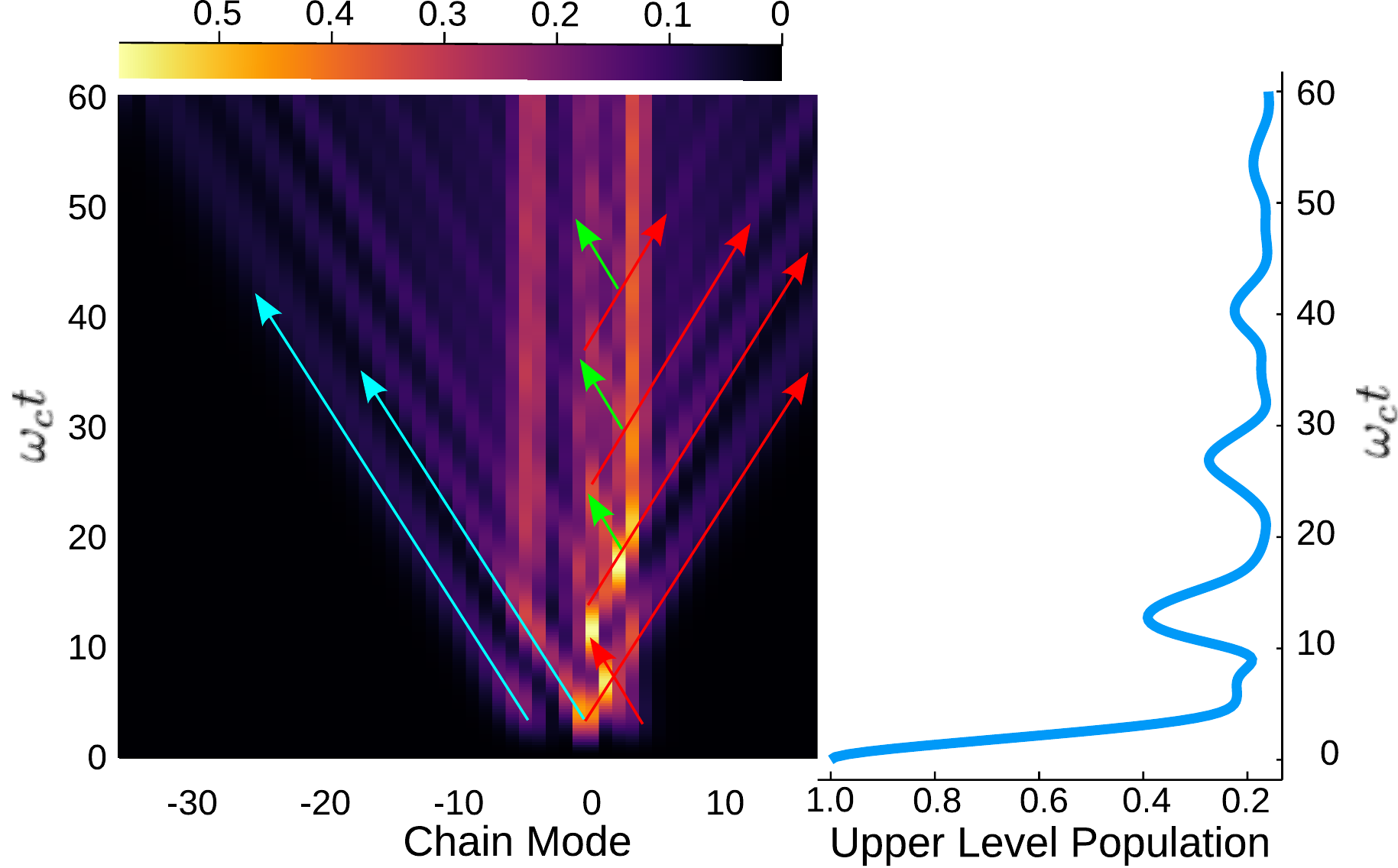}
    \caption{(Left) A heatmap of the chain occupation in time showing the propagation of bath excitations along the chains. Arrows have been added to represent the trajectories of chains' excitations. (The unannotated figure is available in Appendix \ref{sec:unannotated}.) (Right) System eigen-sates population for an initial state in the upper eigenstate). The separation between the two sites is $R = 20$, their coupling is $J = 0.25$, the speed of sound is $c = 2$, $\alpha = 0.12$ and $k_c = 1$. We can definitely see a revival of population at a time consistent with the amount of time needed for a bosonic excitation to travel into the bath from one system's site to the other.}
    \label{fig:Upper_R=20_c=2_T=0}
\end{figure}

Increasing the propagation speed of the bath excitations we can generate several revivals with something like an echo between the two sites, as shown in Fig.~\ref{fig:Upper_R=20_c=2_T=0} where revivals with decreasing amplitudes can be observed with a periodicity of $R/c$.
All the parameters are the same as in Fig.~\ref{fig:Upper_R=40_c=1_T=0} except the speed of the bath's excitations that has been doubled.

The left panel of Fig.~\ref{fig:Upper_R=20_c=2_T=0} shows the heatmap of the chains for the same parameters as Fig.~\ref{fig:Upper_R=40_c=1_T=0} except the speed of bosonic excitation $c$ which is doubled.
We note that even though $c$ is doubled, the speed of the excitation on the chain remains the same as the rays in both figures \ref{fig:Upper_R=40_c=1_T=0} and \ref{fig:Upper_R=20_c=2_T=0} travel the same distance along the chain in the same time.
The propagation speed on the chain is independent of the coupling strength $\alpha$ or the bosonic excitation speed $c$.
The propagation speed on the chain depends on the asymptotic hopping energy between the sites of the chain which depends on the cut-off frequency $\omega_c$ of the spectral density which is here held constant \cite{Chin2010}.
However, for a fixed separation $R$, for $c=2$ the modes for which the coupling between the chain and the second system site is maximal are twice as close to the origin as the ones for $c=1$ (as seen in Sec.~\ref{sec:couplings_T=0}).
Hence, for a given $R$, it takes half the time for an excitation to travel from the second to the first system site for $c=2$ than for $c=1$.
The four revivals of eigen-population that we see in Fig.~\ref{fig:Upper_R=20_c=2_T=0} correspond to the four rays on the positive chain that come from internal reflections of the initial chain excitation highlighted with arrows.
These rays correspond to transmitted parts of bath's excitations bouncing back and forth between the two system sites.

\begin{figure}[h!]
    \centering
    \includegraphics[scale=.4]{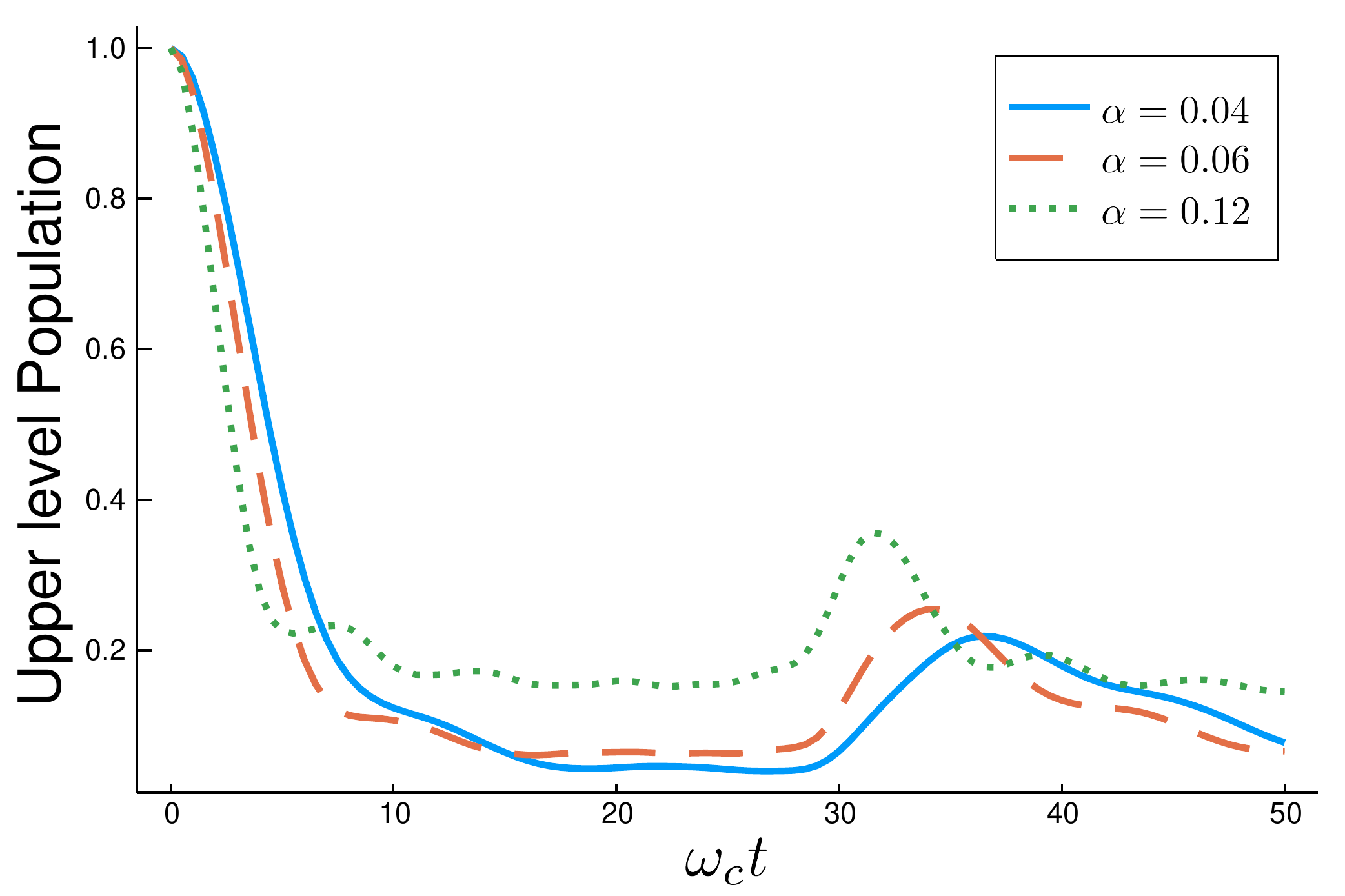}
    \caption{Comparison of the dynamics of the upper eigenstate at zero temperature for different values of the coupling to the bath $\alpha$. As the coupling increases, the revivals become sharper. The other parameters are held constant at $R = 30$, $k_c = 1$, $c =1$ and $J = 0.25$.}
    \label{fig:comparison_alpha_R=30}
\end{figure}

To see the influence of the coupling strength $\alpha$ between the system and the bath, we varied it while keeping a fixed separation $R$ between the system's sites and a fixed speed of the bosonic excitation $c$.
These results are presented in Fig.~\ref{fig:comparison_alpha_R=30} where we can see that increasing the coupling strength sharpens the revivals and brings their peaks closer to $\omega_c t \approx R/c$.
The amplitude of the revivals decrease with the increase of the upper level population prior to the revival.

\begin{figure}[h!]
    \centering
    \includegraphics[scale=.4]{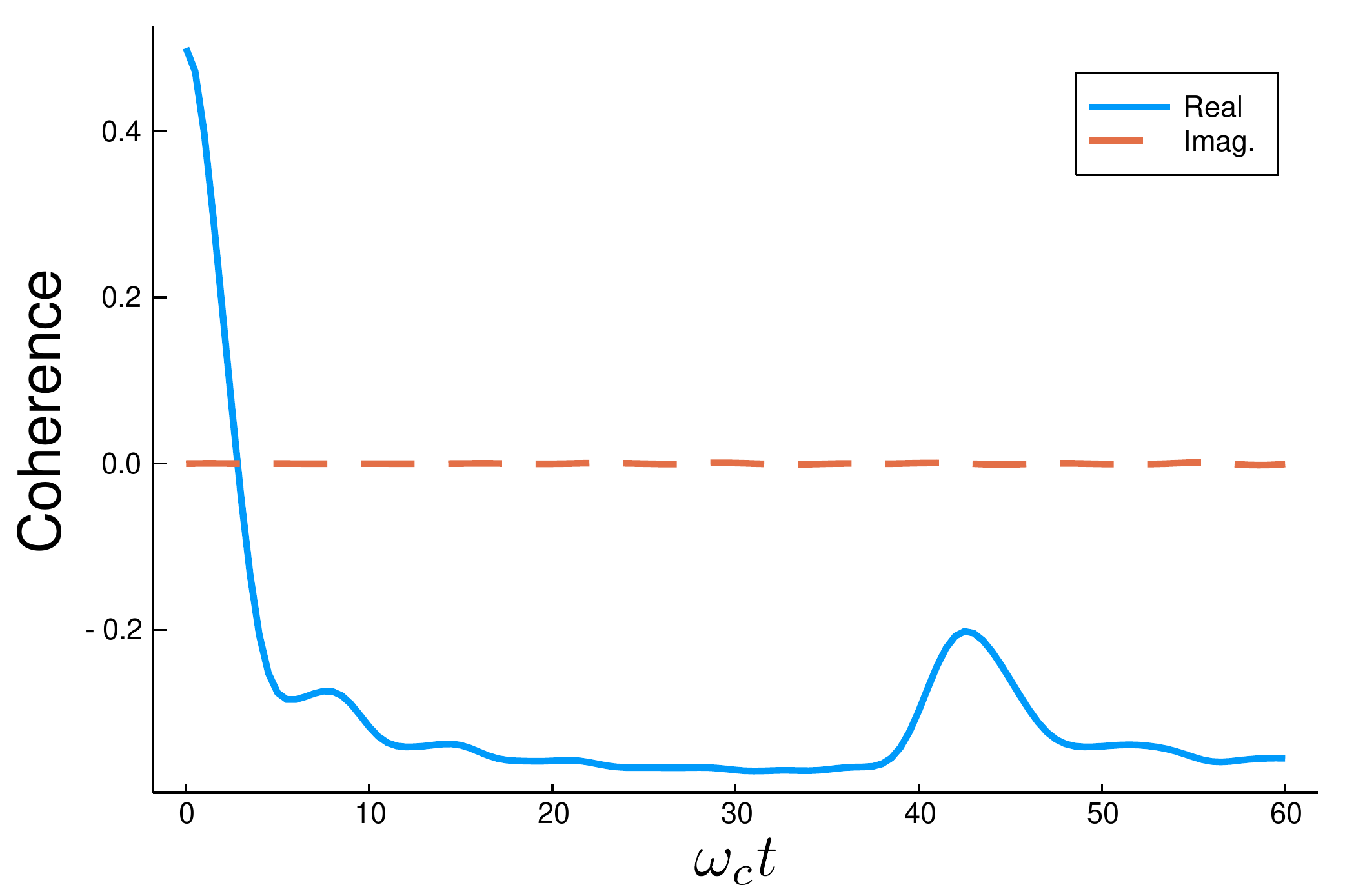}
    \caption{Real and imaginary part of the coherence between the two system sites. The real part is proportional to the upper eigenstate population, hence the revival coincides with a sudden loss of coherence.}
    \label{fig:site_coherence_R=40_T=0}
\end{figure}

Figure \ref{fig:site_coherence_R=40_T=0} shows the coherence between the two sites in the case described by Fig.~\ref{fig:Upper_R=40_c=1_T=0} where the initial state of the system is the upper eigenstate.
For a degenarate TLS, the coherences are proportional to the the upper eigenstate population.
This means that the revivals coincide with a decrease of coherences in absolute value.
A decrease of coherences is hence associated with re-localisation.

\begin{figure}[h!]
    \centering
    \includegraphics[scale=.4]{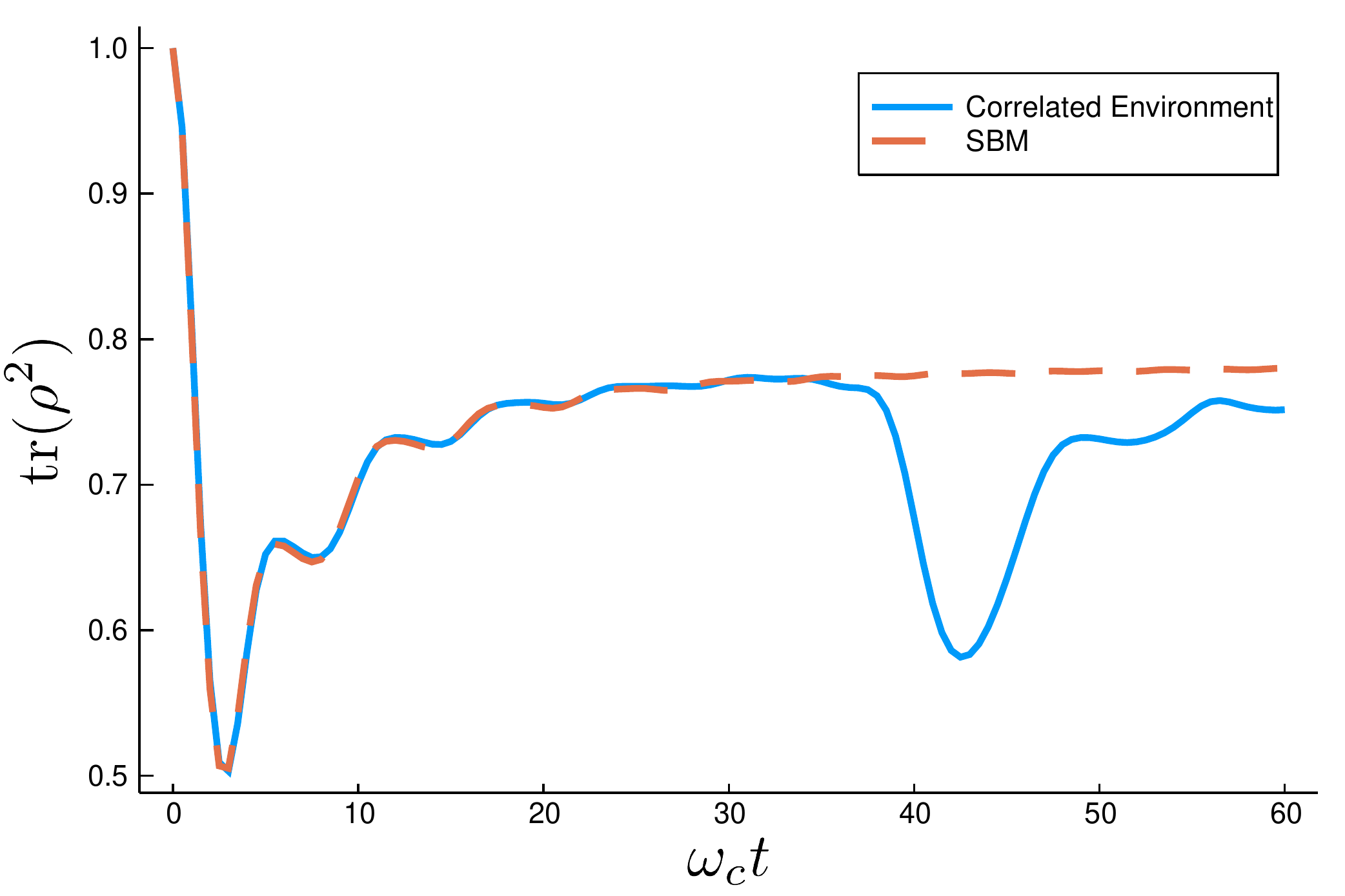}
    \caption{Purity $\tr[\rho_S^2]$ of the system. The revival corresponds to a loss of purity.}
    \label{fig:purity_R=40_T=0}
\end{figure}

Another way to show that this revival of eigen-population (relocalisation) is an incoherent mechanism is to look at the evolution of the purity $\lambda = \tr[\rho_S^2]$ of the system state.
The purity measures how close state is to a pure state:
For $\lambda = 1$, the state is a pure state and for $\lambda = 0.5$ the state of a two level system is a maximal statistical mixture.
Figure \ref{fig:purity_R=40_T=0} presents the evolution of the purity, and clearly shows that revivals are associated with an increase of mixedness of the system's state.

Hence the mechanism behind the revivals can be seen as a partial measurement by the environment on the system's sites that, as a consequence, re-localizes the system's excitation.

\section{Finite Temperature}

\subsection{Couplings}
\begin{figure}[h!]
    \centering
    \includegraphics[scale=.4]{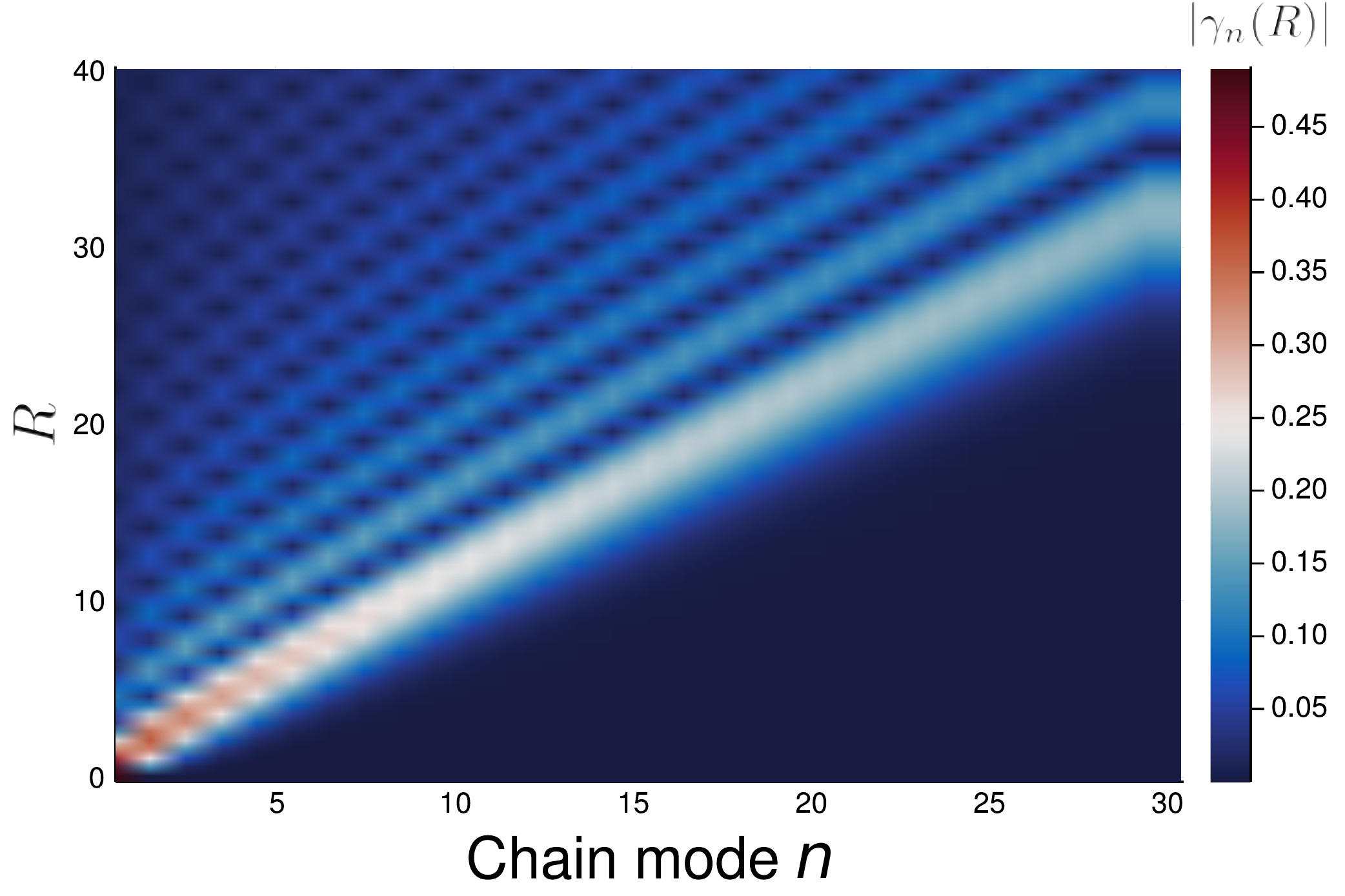}
    \caption{Absolute value of the system-chain coupling constants at finite temperature, for a bosonic bath with a hard cut-off Ohmic spectral density, as a function of the chain modes $n$ and the site separations $R$. The peaks are centered around $n = R/c$. Here $\alpha = 0.12$, $\beta=0.5$, $c = 1$ and $k_c = 1$.}
    \label{fig:abs_couplings_T}
\end{figure}

\begin{figure}
    \centering
    \includegraphics[scale=.4]{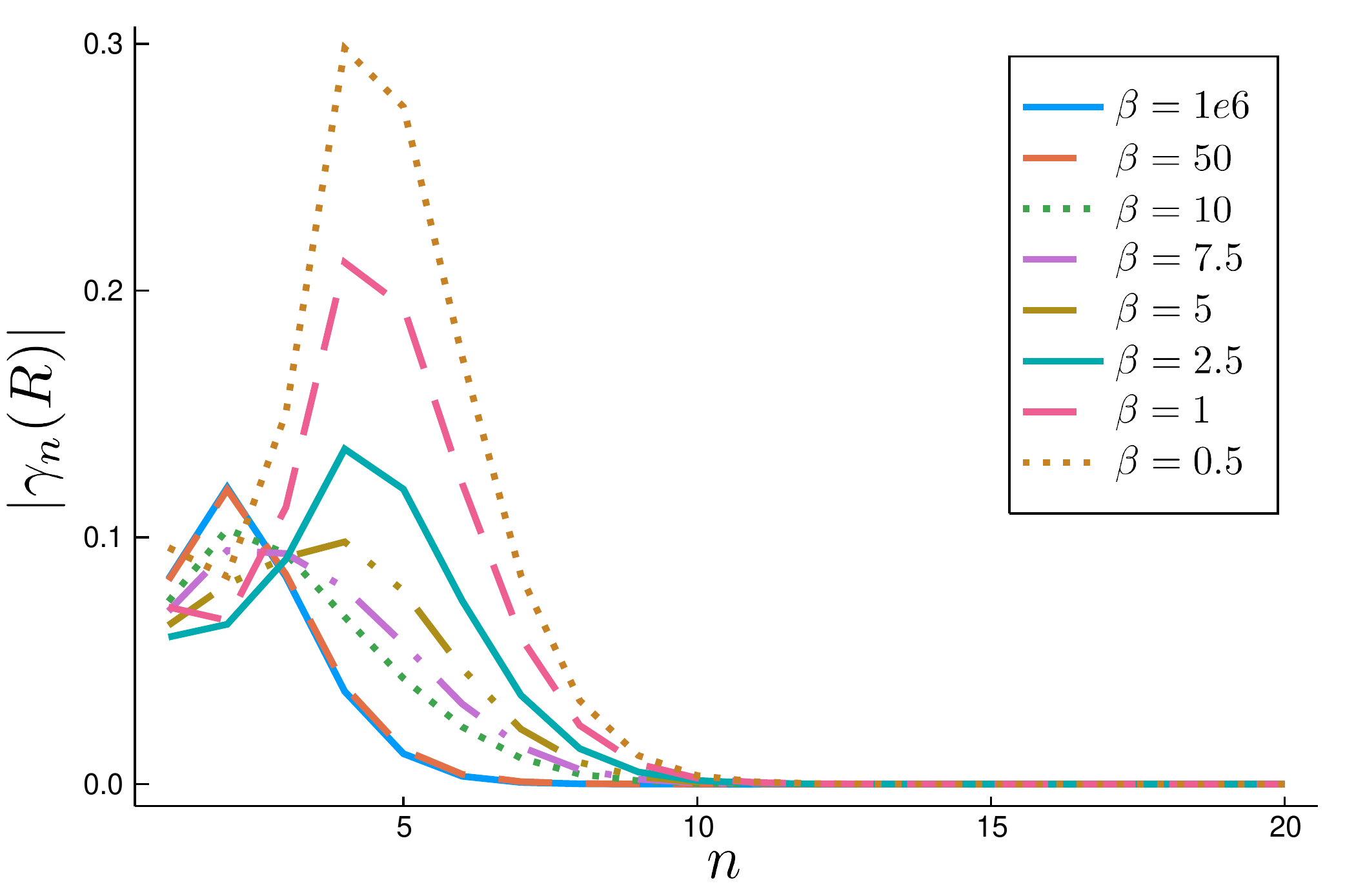}
    \caption{Absolute value of the system-chain coupling constants at finite temperature, for a bosonic bath with a hard cut-off Ohmic spectral density, as a function of the chain mode number $n$ for a fixed $R = 5$ and several temperatures ($\alpha = 0.12$ and $k_c = 1$).}
    \label{fig:abs_couplings_severalT}
\end{figure}

The finite temperature coupling constants between the system and the chain keep broadly the same form as the zero temperature ones.
An example profile for several different system site separations is displayed in Fig.~\ref{fig:abs_couplings_T}.
The differences are that the amplitudes increase with temperature, and the peak value is no longer centered around the mode $n = R/2c$ but rather $n = R/c$.
For $\beta = 0.5$ the amplitude of the coupling is doubled compared to the zero temperature case.
We also note that the tail before the peak presents more oscillations than the zero-temperature one which is smoother.
The change in the coupling profile as a function of temperature is shown in Fig.~\ref{fig:abs_couplings_severalT}.
For high and moderately high temperatures, the couplings decrease in amplitude as $\beta$ increases but are still centered around $n \approx R/c$.
For high values of $\beta$, the amplitude stays constant but the maximum swaps to $n \approx R/2c$ as we recover the zero temperature value.

\subsection{Non-Markovian recurrences and bath feedback}\label{sec:system_beta}

\begin{figure}[h!]
    \centering
    \includegraphics[width=\columnwidth]{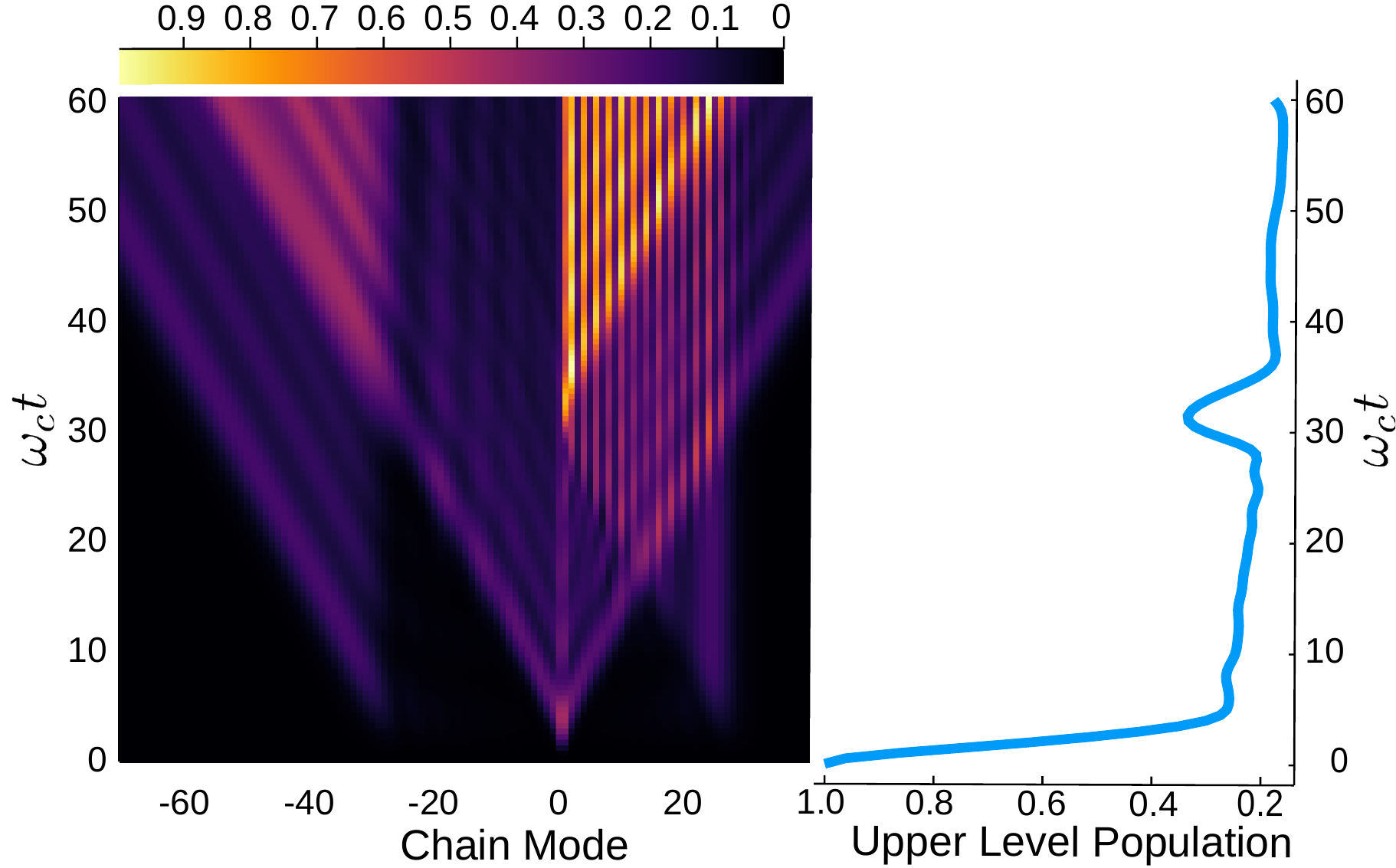}
    \caption{(Left)  A heatmap of the chain occupation in time showing the propagation of bath excitations along the chains. (Right) Upper eigenstate population. The separation between the two sites is $R = 30$, the speed of sound is $c = 1$, the inverse temperature $\beta = 5$ and $\alpha = 0.12$.}
    \label{fig:Pop_and_bath_beta=5}
\end{figure}

Using the method presented in Sec.~\ref{sec:chain_mapping_finiteT}, we also investigated the finite temperature dynamics of the system.
For a large range of values of $\beta$, the system's dynamics stay qualitatively the same except that the steady state population is increased because of thermal fluctuations, as we can see for $\beta = 5$ in Fig.~\ref{fig:Pop_and_bath_beta=5}.

The peak of the coupling is at $n = R/c$ and not $R/2c$ as in the zero temperature case, but the propagation speed along the chain is doubled because the support of the extended spectral density is twice as large as the support of the zero-temperature spectral density \cite{Tamascelli2019}.
The left part of Fig.~\ref{fig:Pop_and_bath_beta=5} shows the time-frequency diagram for finite temperature for the inverse temperature $\beta = 5$ and a separation $R=30$.
For this intermediate temperature, the chain excitation propagate balistically in way similar to the zero temperature case, except that modes are more populated thanks to thermal fluctuations.
Wave-packets emitted from the origin of the chain and the part coupled to the second site interfere when they meet.
Hence, we see interference fringes appear when excitations with different phases come together.
As in the finite temperature case, when excitations reach the origin of the chain they give rise to a revival of the eigenstate population.

\begin{figure}[h!]
    \centering
    \includegraphics[scale=.4]{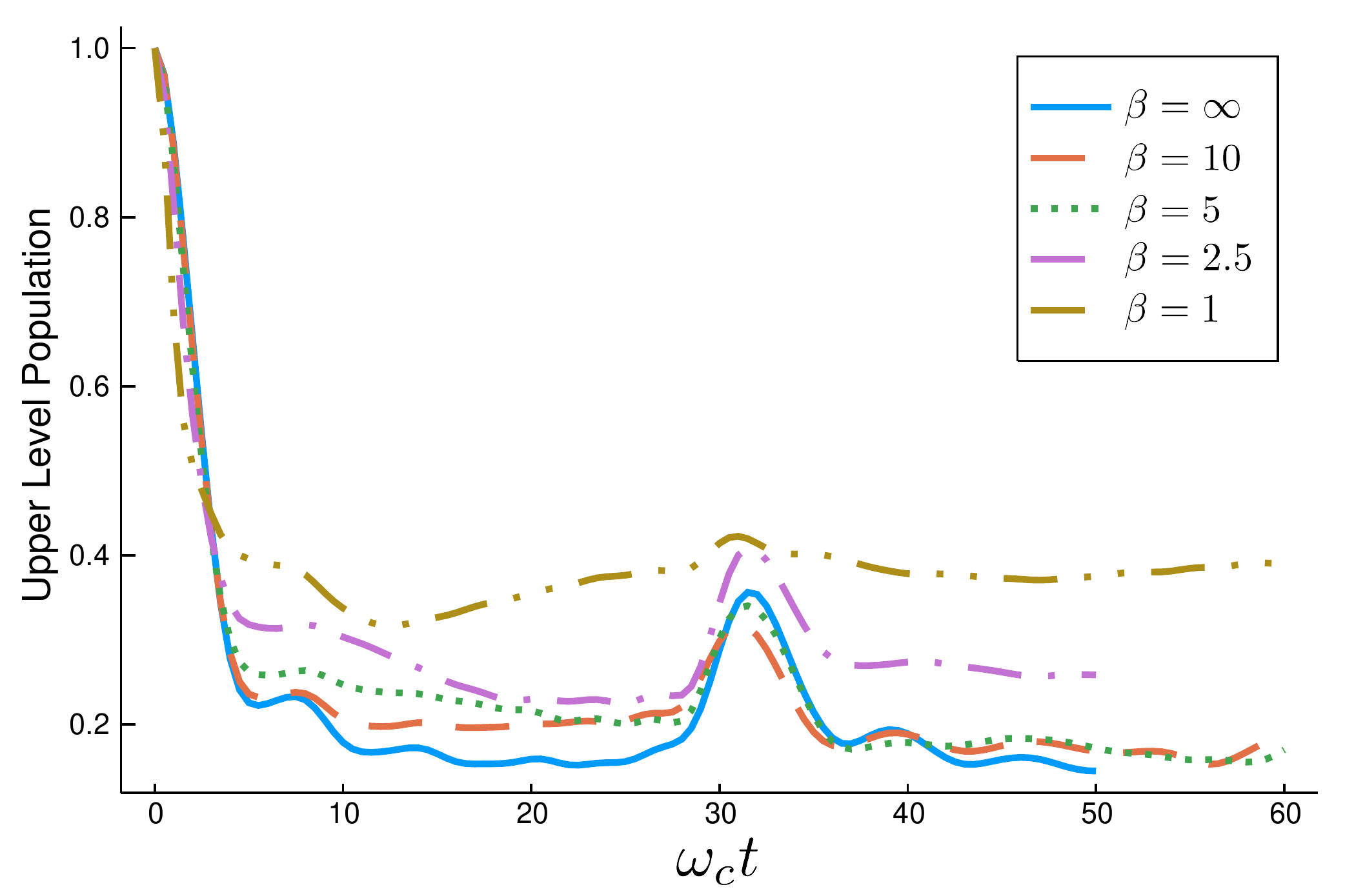}
    \caption{Upper eigenstate populations for $R = 30$, $\omega_c = 1$, $c =1$, $\omega_0 = 0.25$ and $\alpha = 0.12$ for several values of the inverse temperature $\beta$. }
    \label{fig:Pop_several_beta}
\end{figure}

\begin{figure}
    \centering
    \includegraphics[width=\columnwidth]{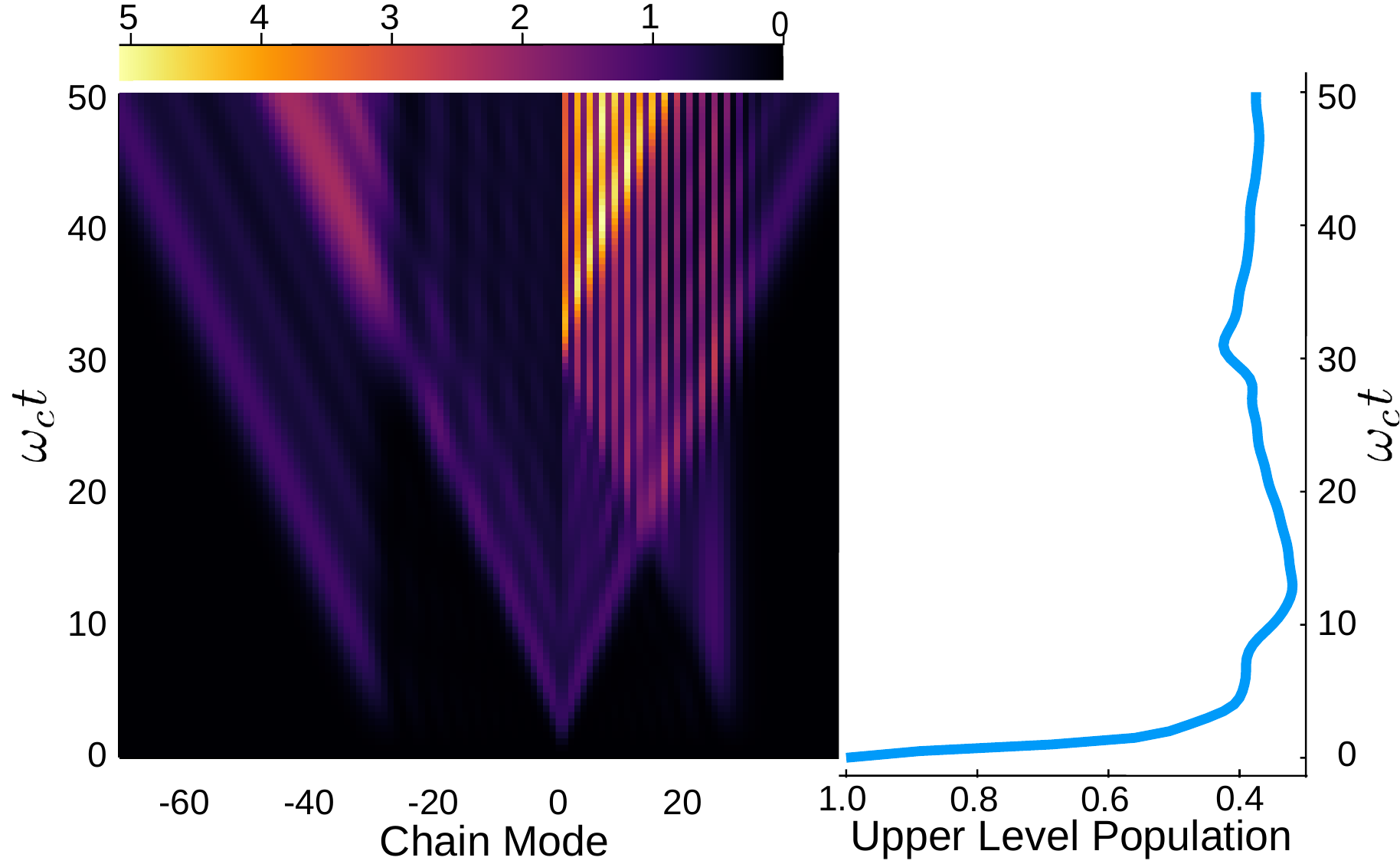}
    \caption{(Left)  A heatmap of the chain occupation in time showing the propagation of bath excitations along the chains. (Right) Upper eigenstate population. For high-temperature the revival is less pronounced. The separation between the two sites is $R = 30$, the speed of sound is $c = 1$, the inverse temperature $\beta = 1$ and $\alpha = 0.12$.}
    \label{fig:Pop_and_bath_beta=1}
\end{figure}

Figure \ref{fig:Pop_several_beta} shows the upper eigenstate population for increasing values of the temperature.
The revivals are still present for moderate temperatures such as $\beta = 5$ but they become barely noticeable for high-temperature, as we can also see in Fig.~\ref{fig:Pop_and_bath_beta=1}.
Between $\beta = 5$ and $\beta = 1$ the dynamics of the chains' modes are the same but the populations are increased by a factor $\sim 5$.
This increased population is a direct consequence of the thermal population.
We can see, in Fig.~\ref{fig:Pop_several_beta}, that the amplitude of the revival seems to be related to the depth of the plateau reached before $\omega_c t \approx R/c$.
Hence, as the eigen population in this region gets closer to a half, the revival is suppressed.

For higher temperature, as in Fig.~\ref{fig:Pop_and_bath_beta=1}, the behaviour of the chain is akin to the one we could see for a SBM with a Ohmic spectral density \cite{Tamascelli2020} but duplicated on the chain.
As they propagate on the chain, excitations leave a trail of populated modes behind them that correspond to the cones we can see on the figure.

\section{Conclusion}
Motivated by the ability of biological nanostructures to coordinate (opto)electronic processes through the relaying of environmental (structural) `signal' motions, we have presented a numerically exact exploration of a model that can describe these highly non-Markovian effects. To do so, we have extended the standard T-TEDOPA techniques, in the 1TDVP formulation, to treat the long-range chain couplings that encode information about spatial correlations. In doing so, we have proved that for system-bath problems with spatially correlated interactions, the Hamiltonian matrix product operator will always have a bond dimension proportional to the number of system states, regardless of the range of the interactions. Provided that  -- as in most models of open systems -- the environment is non-interacting, this allows tensor network to be a computationally powerful method for exploring multisite dynamics where non-Markovian  environmental feedback could lead to functionally relevant non-equilibirum states and/or transient effects that could materially alter the outcome of a process, if a certain set of events precede it.              

As our first exploration of this aspect of highly structured nanoscale dissipation, we have shown that one of the simplest conceptual forms of correlated environments (plane waves in 1D) supports strong spatio-temporal feedback effects that introduce new timescales into the dissipative dynamics and show clear signs of having stored information about the early time motion, i.e. after sharp decays, we find sharp revivals. Moreover, we have also found that periodic behaviour with $T=R/c$ can also be obtained in which each revival acts as a generator of subsequent revivals, leading to periodic -- but highly anharmonic -- energy exchange between the system states. Finally, we have shown that finite temperatures tend to broaden and suppress these revival effects, although they visibly persist for temperatures up to the system energy gap.        

These results encouragingly point to the idea that suitably tailored environments could be coupled to electronic processes in order to produce well-defined functional effects at later times and in distant places in the structure. To explore this in more detail requires the inclusion of larger, multi-component systems, and this is something we have shown could be done with the present method. However, in the majority of nanostructures, biological or otherwise, the 1D plane wave environment is likely to be an oversimplification. It be therefore be of future interest to consider different kinds of relationship between mode frequencies and spatial correlation in the system-bath interactions, such as those that can be extracted by molecular dynamics simulations of proteins \cite{olbrich2011quest,zuehlsdorff2021vibronic}, normal mode analysis\cite{renger2012normal,morgan2016nonlinear}, or coarse-grained methods that access the slow, large amplitude motions of complex structures \cite{chaillet2020static,fokas2017evidence}. Given that the present method works with arbitrarily structured spectral functions and can handle long-range system-environment interactions in the chain or tree tensor representations of the problem, we hope that this work will encourage further examination of the no-doubt rich functional phenomenology of spatially correlated open quantum systems.   

\acknowledgments

TL, AWC and BWL thank the Defence Science and Technology Laboratory (dstl) and Direction G\'en\'erale de l’Armement (DGA) for support through the Anglo-French PhD scheme. 
AD acknowledges support by the École Doctorale 564 Physique en Île-de-France.
DG acknowledges studentship funding from EPSRC (EP/L015110/1).

\onecolumngrid
\appendix

\section{Mapping to SBM for $N=2$}\label{sec:SBMmapping}
The $N=2$ case is a specific case where, because of the symmetry around the mid-point between the two sites, the problem presented in this paper can be written in the form of a SBM with an effective spectral density that depends explicitly on the sites separation.
Consider the interaction Hamiltonian $\hint$ in Eq.~(\ref{eq:hamiltonian2}) in the case of a two-site system with intersite distance $R$, we have
\begin{align}
    \hint &= \sum_{\alpha}\proj{\alpha}\int_{-k_c}^{+k_c}(g_k^{\alpha}\ak + g_k^{\alpha *}\akd)\d k \\
    &= \sum_{\alpha}\proj{\alpha}\int_{0}^{+k_c}\left(g_k^{\alpha}(\ak + \ad_{-k}) + g_k^{\alpha *}(\akd + \a_{-k})\right)\d k \ .
\end{align}
We can introduce a new set of vibrational modes, the symmetric mode $\hat{c}_k$ and the antisymmetric mode $\hat{d}_k$
\begin{align}
    \hat{c}_k &= \frac{\ak + \a_{-k}}{\sqrt{2}}\ , \\
    \hat{d}_k &= \frac{\ak - \a_{-k}}{\sqrt{2}}\ .
\end{align}
Hence, the interaction Hamiltonian becomes
\begin{align}
    \hint = \sum_{\alpha}\proj{\alpha} \int_{0}^{+k_c} \big [ & g_k^\alpha\sqrt{2} (\hat{c}_k + \hat{c}_k^\dagger) + g_k^{\alpha*}\sqrt{2}(\hat{c}_k + \hat{c}_k^\dagger)\nonumber \\
    &+ g_k^\alpha\sqrt{2} (\hat{d}_k - \hat{d}_k^\dagger) -  g_k^{\alpha*}\sqrt{2}(\hat{d}_k - \hat{d}_k^\dagger)\big]\d k
\end{align}
We choose the origin of position at the midpoint between the two sites so that
\begin{align}
    \hint = &\Big(\proj{-R/2} + \proj{R/2}\Big)\nonumber\int_{0}^{+k_c} 2\sqrt{2}g_k \cos\left(\frac{kR}{2}\right)(\hat{c}_k + \hat{c}_k^\dagger)\d k \nonumber\\
    &+ \Big(\proj{-R/2} - \proj{R/2}\Big)\int_{0}^{+k_c} 2\sqrt{2}\i g_k\sin\left(\frac{kR}{2}\right)(\hat{d}_k^\dagger - \hat{d}_k) \d k\\
    \hint = & ~\id_S~\mathrm{const} + \hat{\sigma}_z\int_{0}^{+k_c} 2\sqrt{2}\i g_k\sin\left(\frac{kR}{2}\right)(\hat{d}_k^\dagger - \hat{d}_k) \d k
\end{align}
Therefore the system only couples to the antisymmetric vibration modes and thus corresponds to a SBM with an effective spectral density $J_\text{eff}(k) = 8|g_k|^2\sin^2(\frac{kR}{2})$.
However for larger values of $N$ it is no longer possible to map the system to a SBM.
This is similar to the spin-mapping presented in \cite{strathearn}.

\section{Limit Cases}
\label{sec:convergence}
\subsection{Large Separation}
\begin{figure}[h]
    \centering
    \includegraphics[width=0.5\columnwidth]{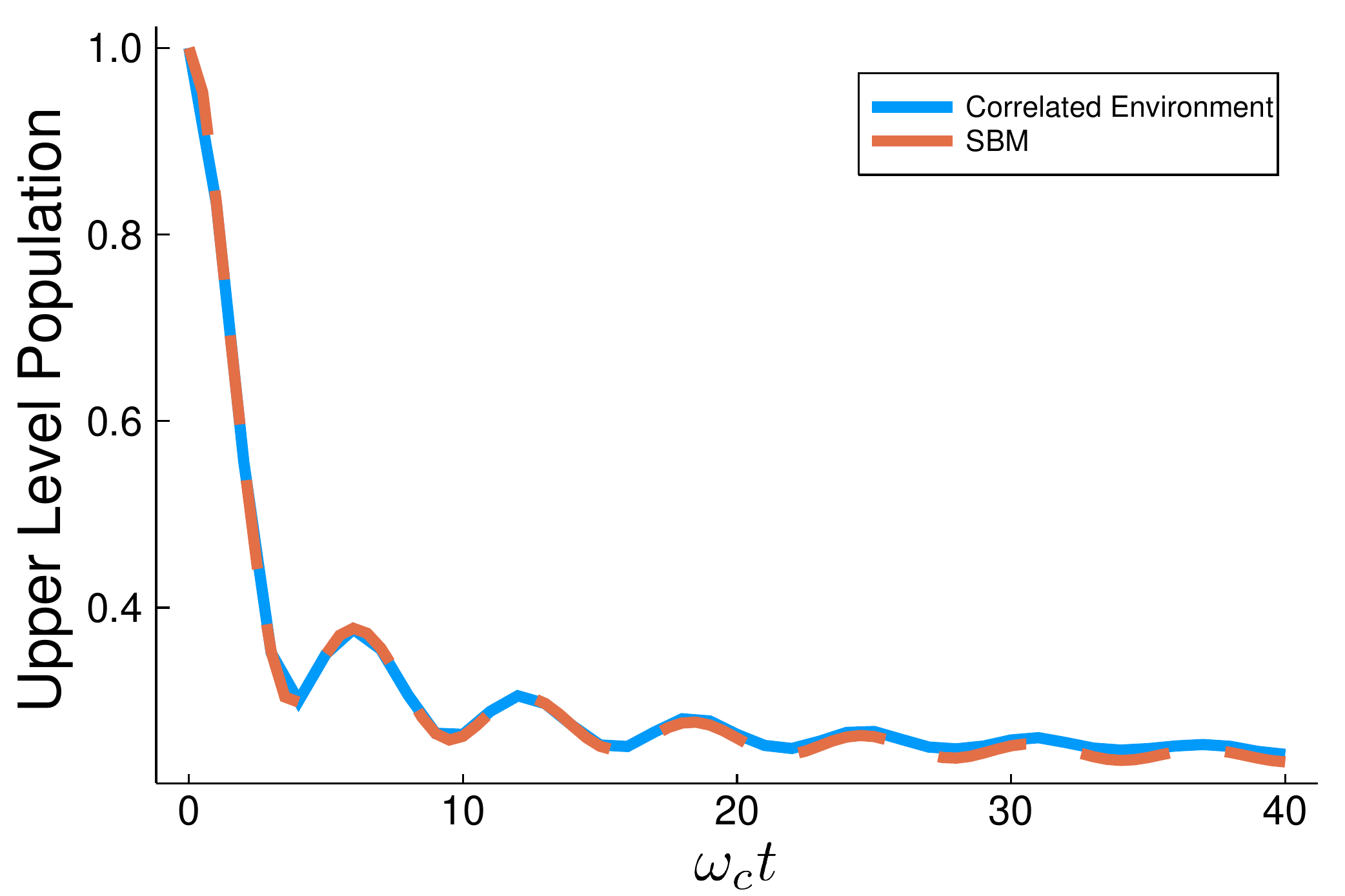}
    \caption{Dynamics of the up-state $\ket{\uparrow_z}$ of a Spin Boson Model (SBM) compared with the dynamics of the upper eigenstate of the Correlated Environment model for corresponding parameters ($k_c = 1$, $c =1$, $J = 0.25$ and $\alpha = 0.2$) with a large separation $R=200$ between the two sites of the system. The two dynamics are the same.}
    \label{fig:RtoInfty_vs_SBM}
\end{figure}

According to the coupling structure presented in Sec.~\ref{sec:couplings_T=0}, the further away the two sites of the system are, the less the second site interacts with the beginning of the chain.
Thus, we can expect that for infinite separation when $R\to\infty$ this system will behave like a SBM.
Figure \ref{fig:RtoInfty_vs_SBM} shows the comparison between the SBM and the infinite separation case.

\subsection{Low Temperature}
\begin{figure}
    \centering
    \includegraphics[width=0.5\columnwidth]{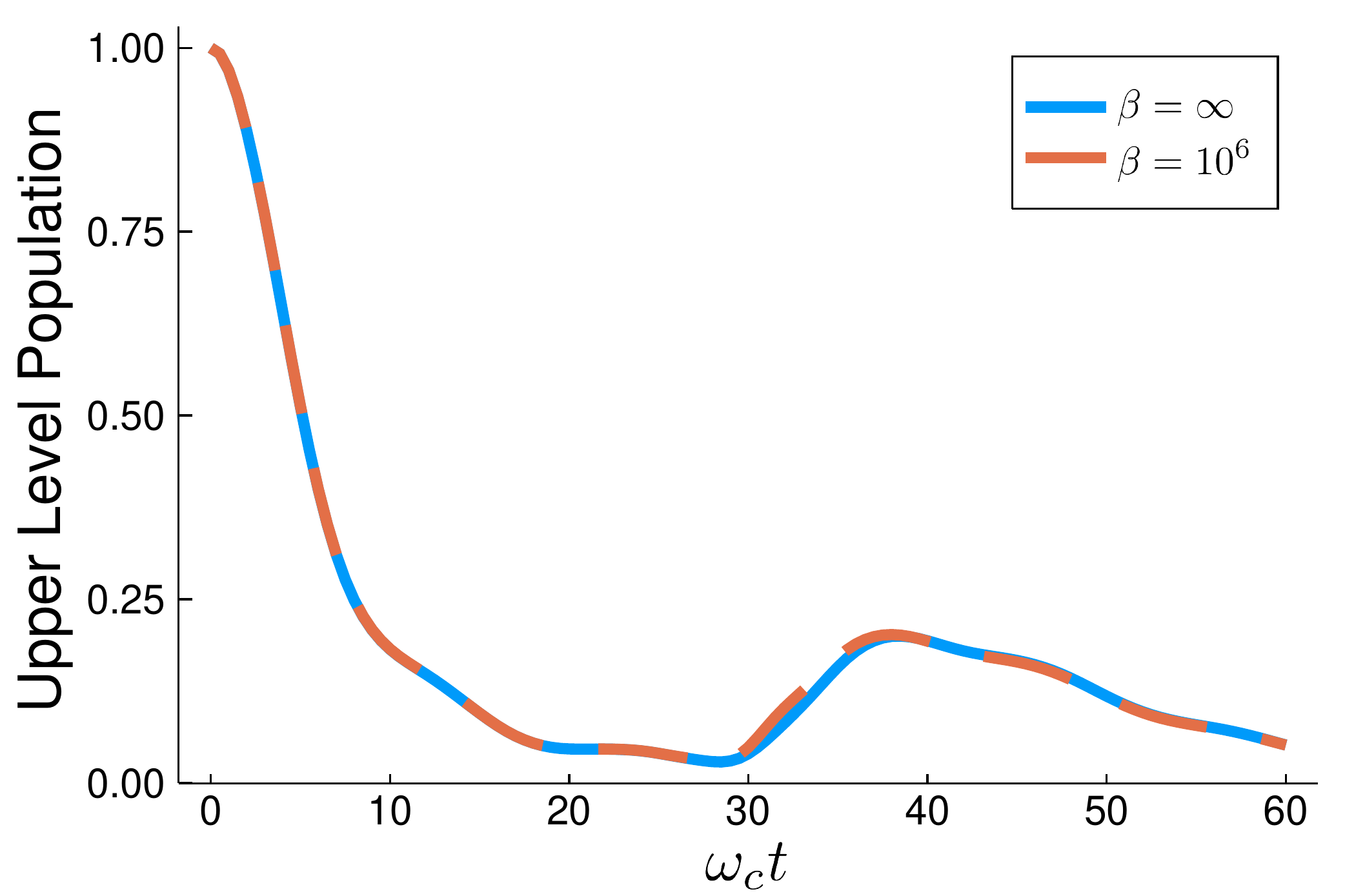}
    \caption{Upper Level Population for zero temperature (solid line) and $\beta = 10^6$ (dashed line) all other parameters being the same ($\alpha = 0.03$, $J=0.25$, $c=1$ and $k_c =1$). The dynamics obtained with the zero-temperature and finite-temperature algorithms are identical.}
    \label{fig:LowTemperatureConvergence}
\end{figure}

The finite temperature effective spectral density $J_\beta(k)$ converges toward the zero temperature one when $\beta \to \infty$ as its value for negative wave-vector becomes uniformly null.
Hence, the quantities calculated using this finite temperature function should all converge toward their zero-temperature counterparts when $\beta$ is increased.
Figure \ref{fig:LowTemperatureConvergence} shows that the population dynamics of the zero-temperature case is recovered.

\begin{figure}
\centering
    \subfigure[]{\label{fig:re_couplings_T}\includegraphics[width=0.45\textwidth]{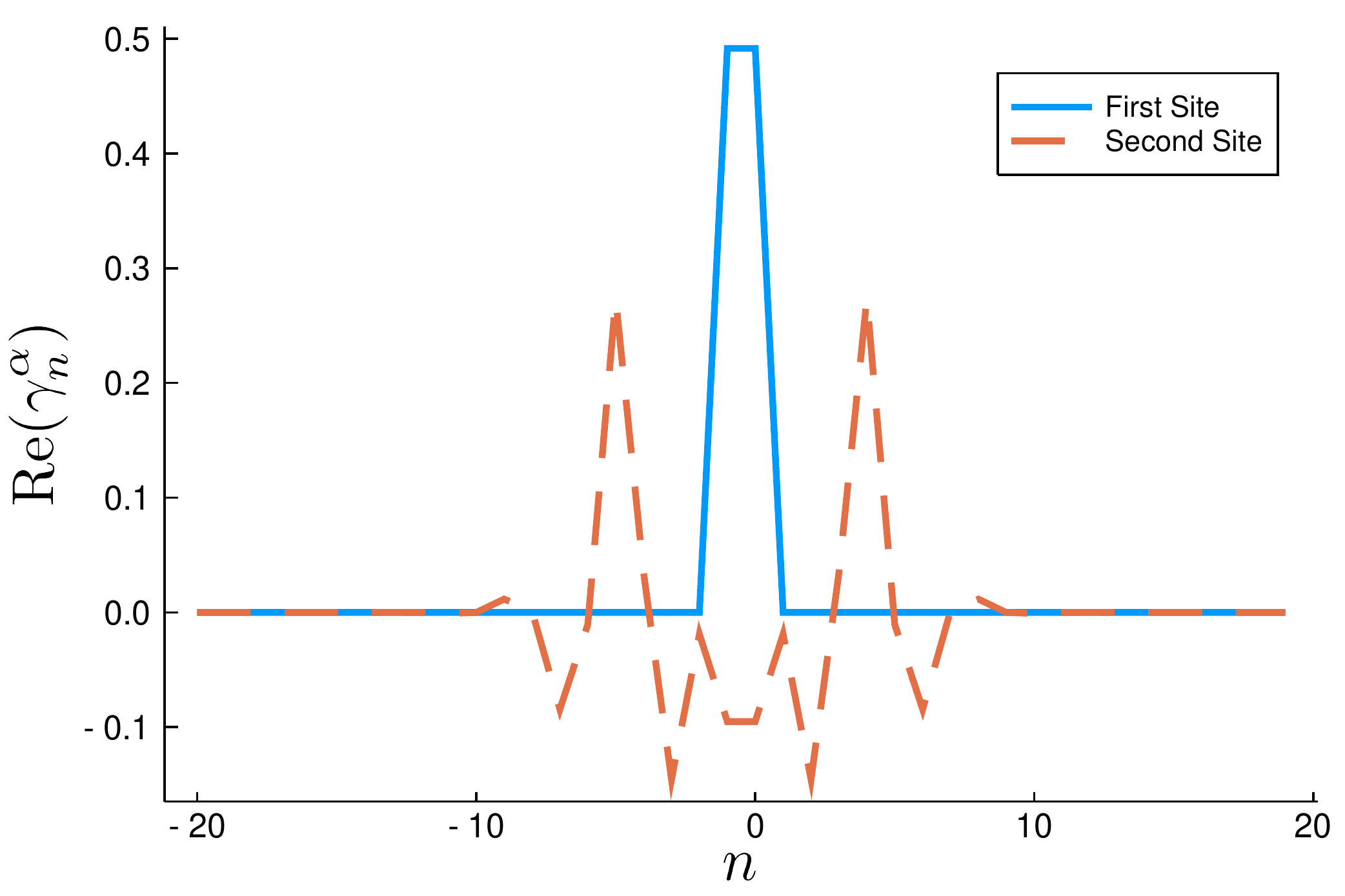}}
    \subfigure[]{\label{fig:im_couplings_T}\includegraphics[width=0.45\textwidth]{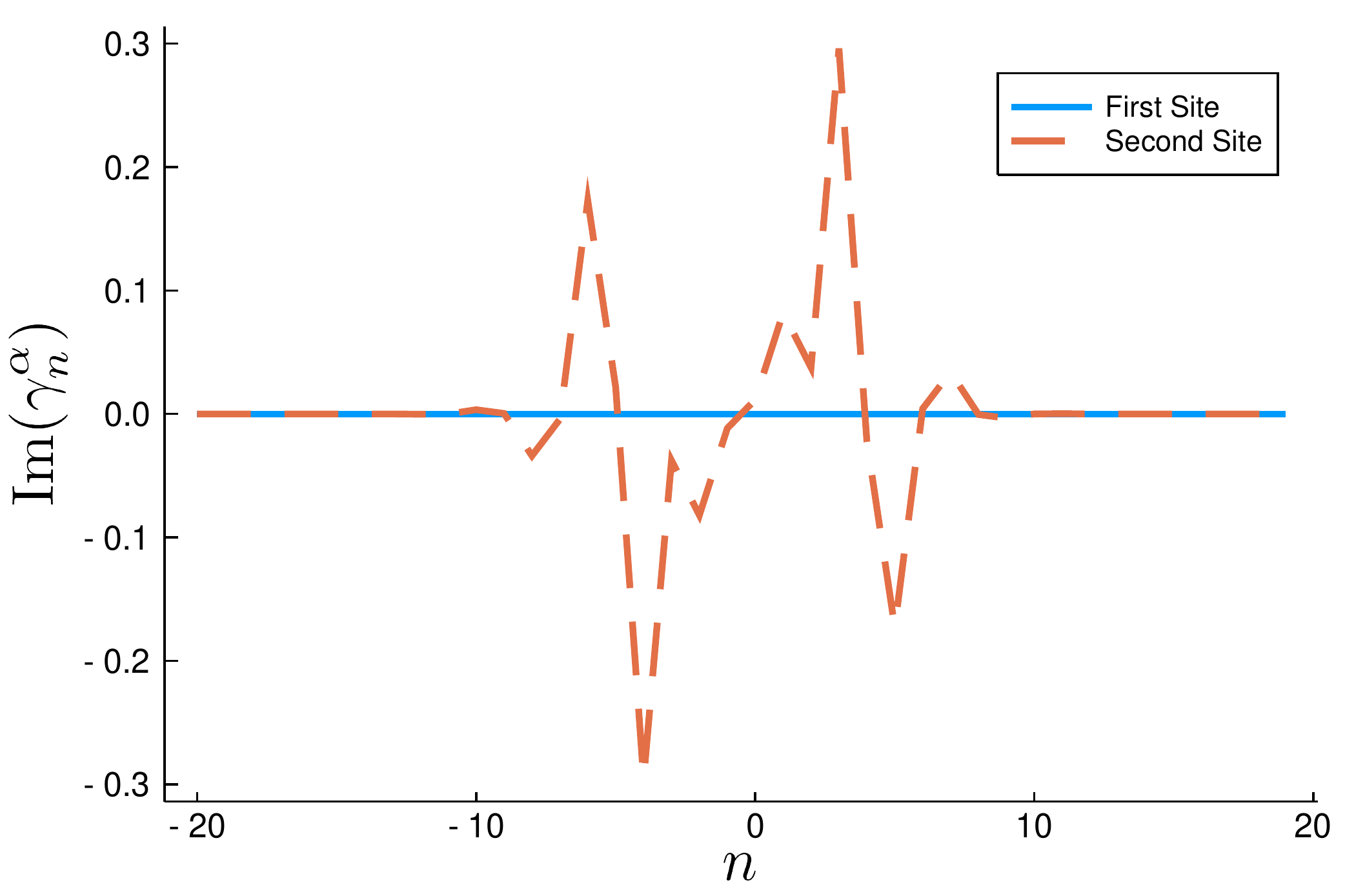}}
\caption{\label{fig:re_im_couplings_T}(a) Real part of the finite-temperature couplings for the two sites of the system as a function of the chain modes.
(b) Imaginary part of the finite-temperature couplings. The negative values along the $x$-axis correspond to the chain of negative wave-vectors and the positive values to positive wave-vectors. Parameters are the same as in Fig.~\ref{fig:abs_couplings_T}, and in particular $\beta = 0.5$.}
\end{figure}

For the same reasons the couplings $\gamma_n(R)$ determined with the finite temperature spectral density should become identical to the zero-temperature ones calculated with the spectral density $J(k)$.
This was already shown in Sec.~\ref{sec:system_beta} with the absolute values of the coupling constants $\gamma_n(R)$ as shown in Fig.~\ref{fig:abs_couplings_severalT}.
The real and imaginary parts of the zero-temperature coupling constants for the two sites are presented in Fig.~\ref{fig:re_im_couplings_T}.
Figure \ref{fig:re_im_couplings_T} shows the coupling constants at a finite temperature $\beta = 0.5$ for comparison.
The finite temperature couplings for a large $\beta$ are presented in Fig.~\ref{fig:re_im_couplings_lowT} and show that the finite temperature coupling coefficients converge to the zero-temperature ones when the limit $\beta \to \infty$ is taken.
\begin{figure}
\centering
    \subfigure[]{\label{fig:re_couplings}\includegraphics[width=0.45\textwidth]{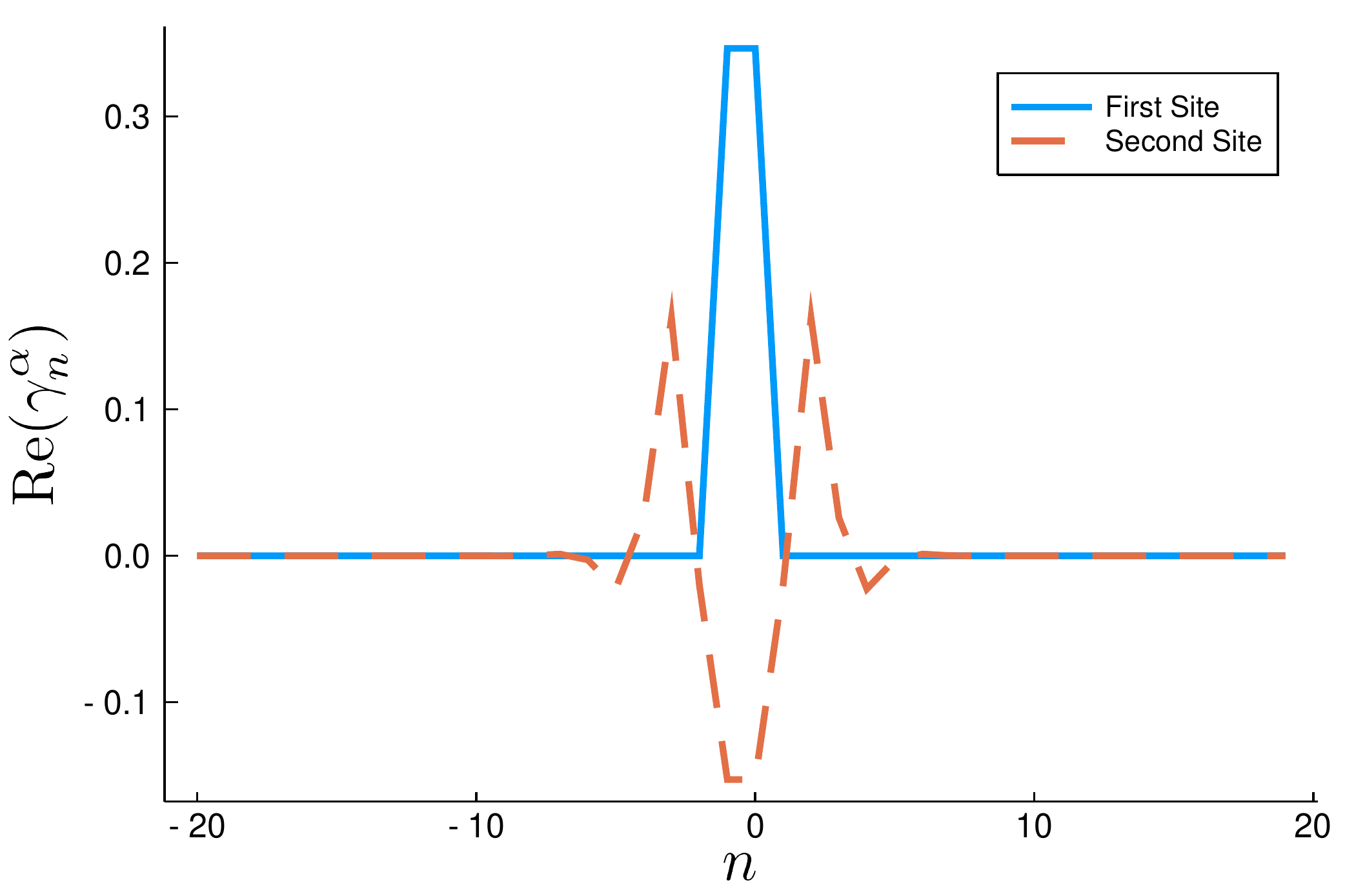}}
    \subfigure[]{\label{fig:im_couplings}\includegraphics[width=0.45\textwidth]{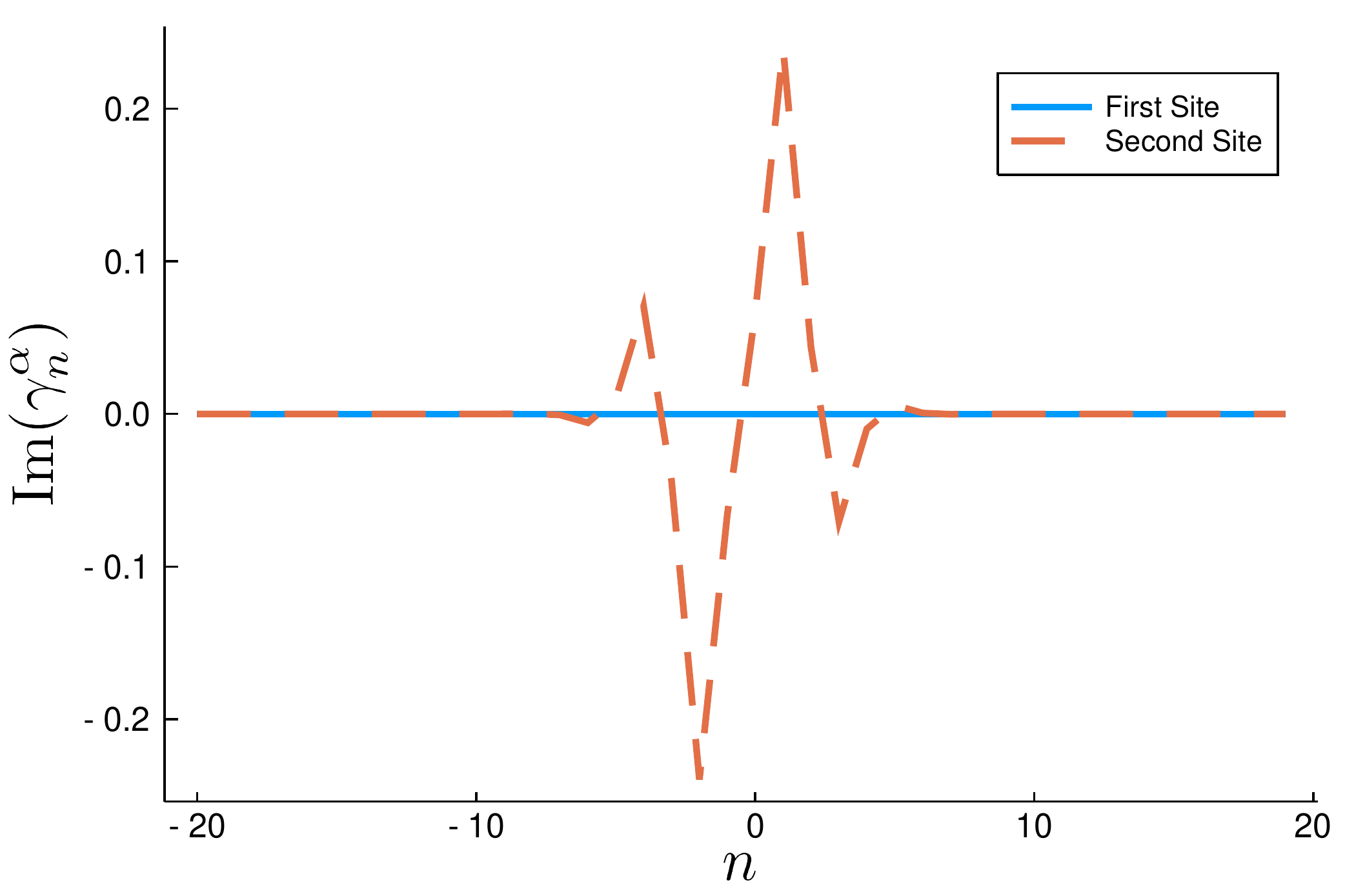}}
\caption{\label{fig:re_im_couplings}(a) Real part of the zero-temperature couplings for the two sites of the system as a function of the chain modes.
(b) Imaginary part of the zero-temperature couplings. The negative values along the $x$-axis correspond to the chain of negative wave-vectors and the positive values to positive wave-vectors.}
\end{figure}

\begin{figure}
\centering
    \subfigure[]{\label{fig:re_couplings_lowT}\includegraphics[width=0.45\textwidth]{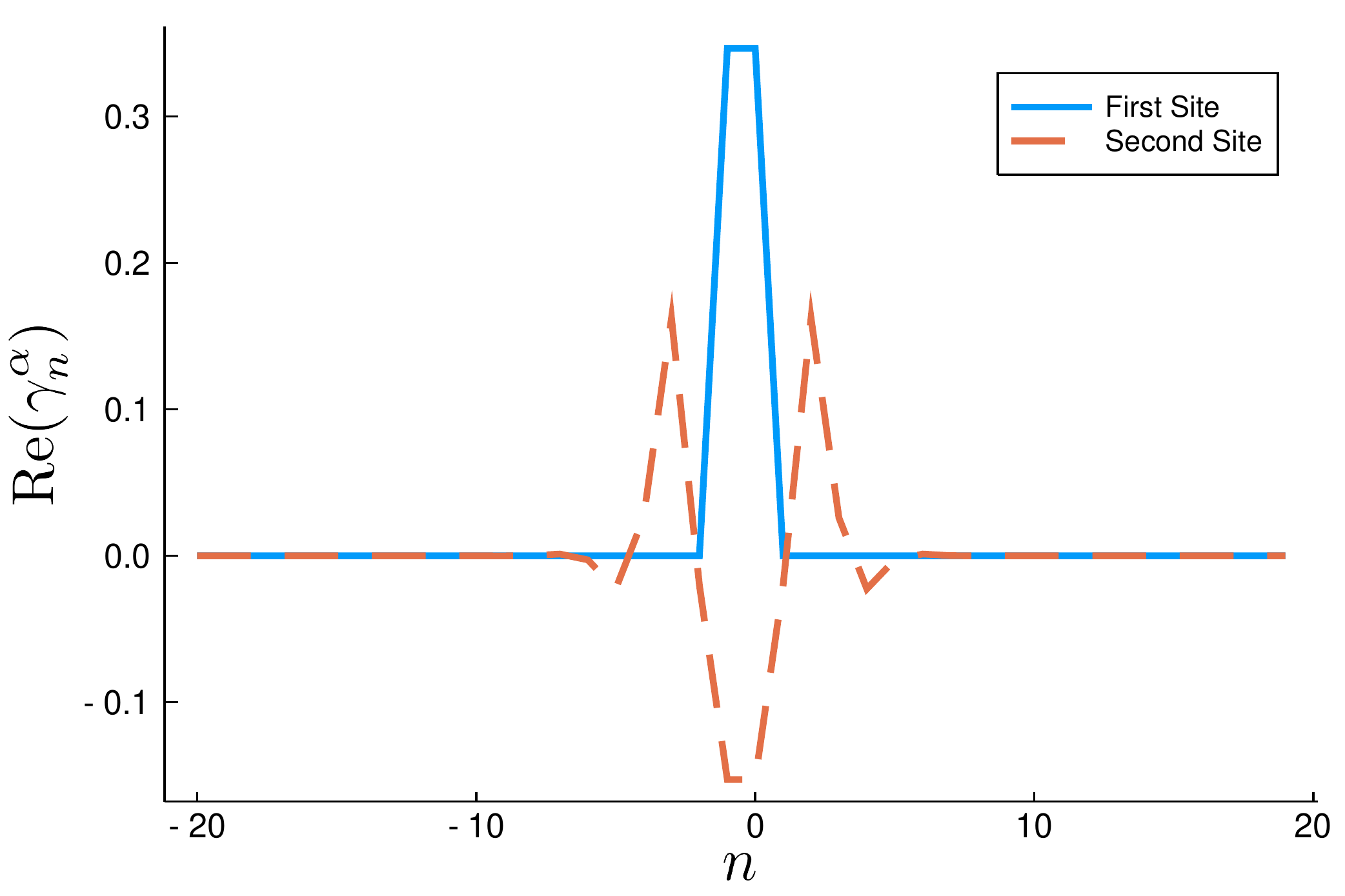}}
    \subfigure[]{\label{fig:im_couplings_lowT}\includegraphics[width=0.45\textwidth]{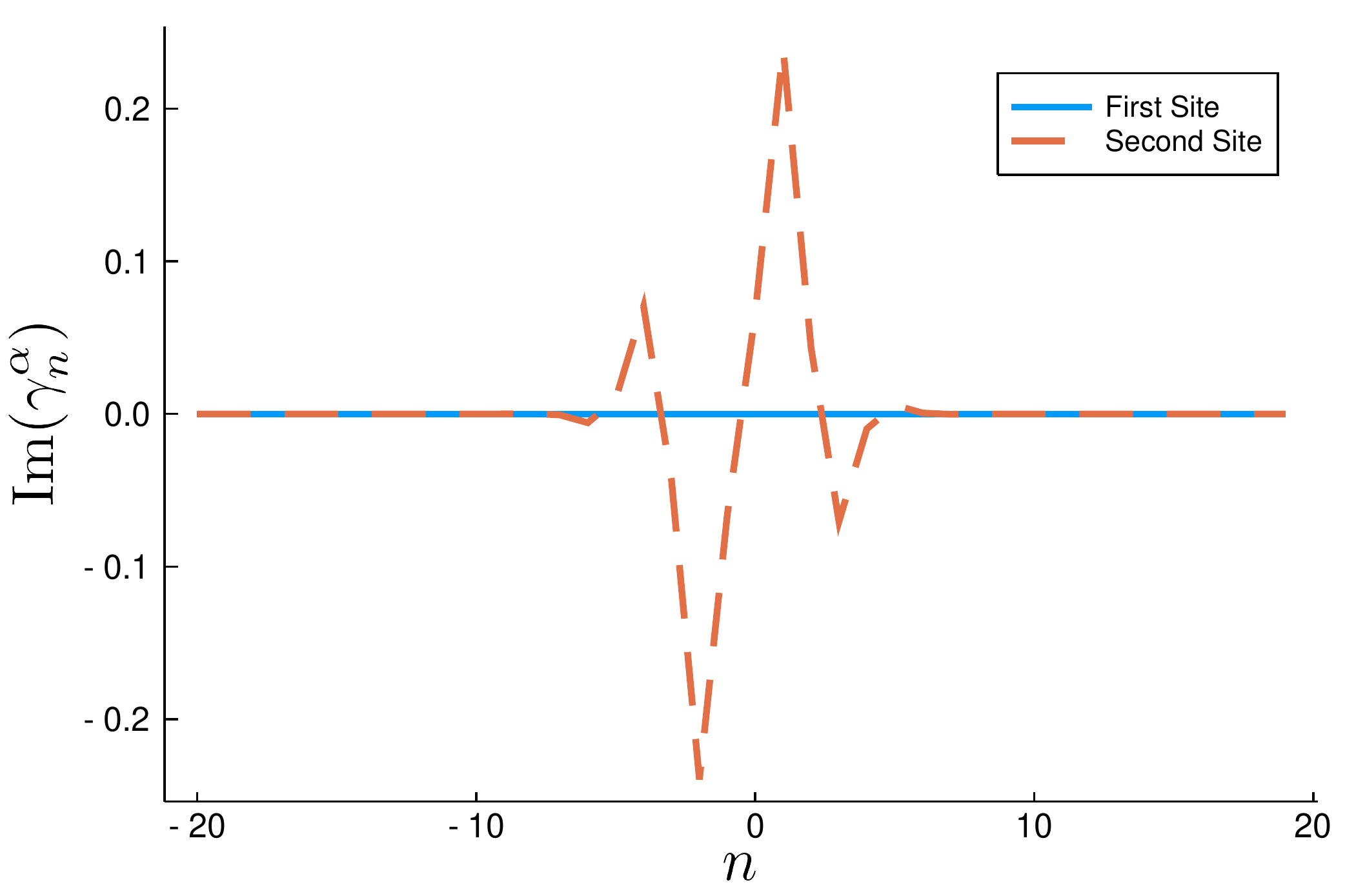}}
\caption{\label{fig:re_im_couplings_lowT}(a) Real part of the finite-temperature couplings for the two sites of the system as a function of the chain modes at $\beta = 10^6$.
(b) Imaginary part of the zero-temperature couplings at $\beta = 10^6$. The behaviour is identical to the one obtained with the zero-temperature algorithm. The negative values along the $x$-axis correspond to the chain of negative wave-vectors and the positive values to positive wave-vectors. The other parameters are the same as in Fig.~\ref{fig:re_im_couplings}.}
\end{figure}

\section{Bath Dynamics}
\label{sec:unannotated}
\begin{figure}
    \centering
    \includegraphics[width=0.5\linewidth]{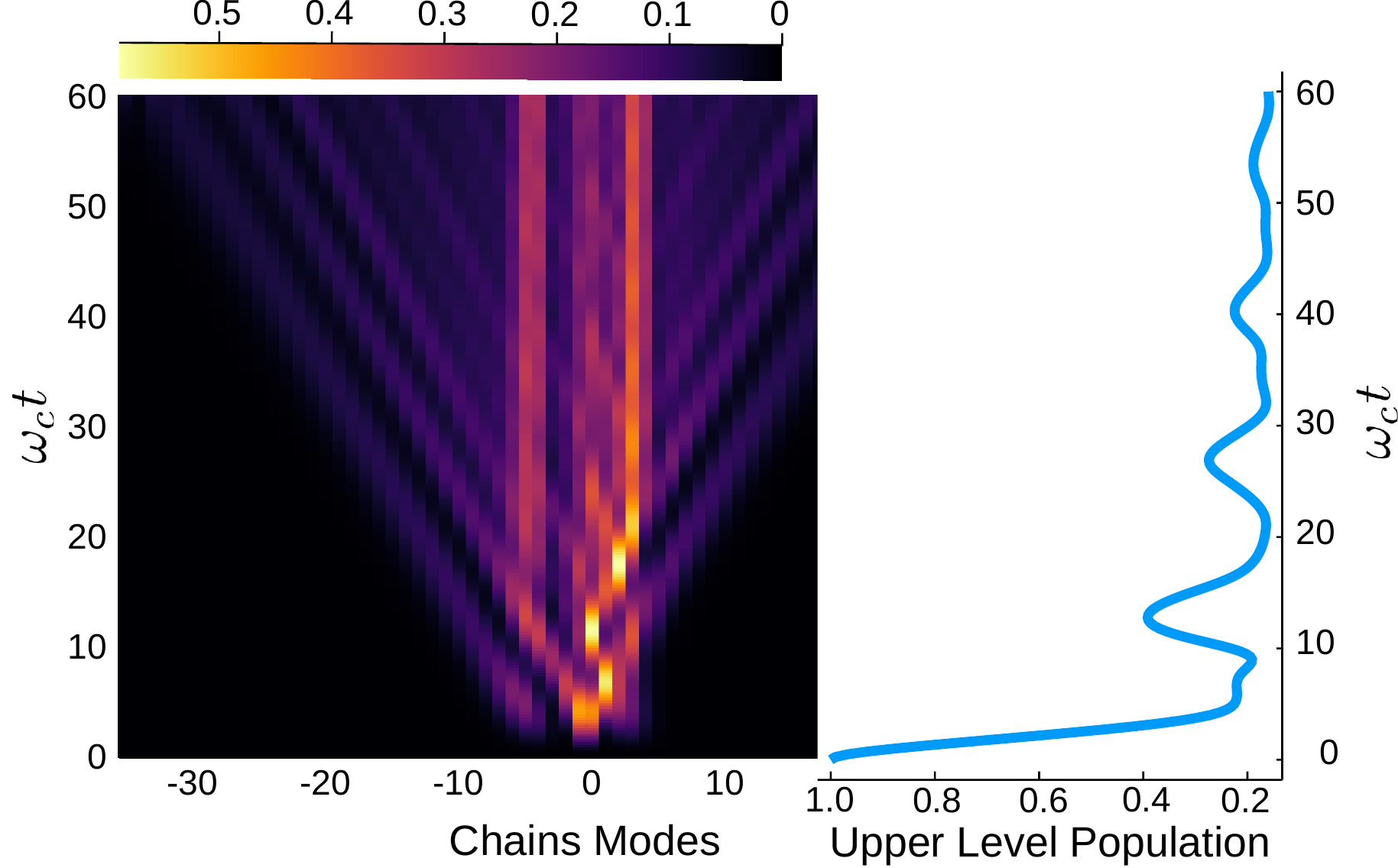}
    \caption{(Left)A heatmap showing the propagation of bath excitations along the chains. (Right) System eigen-sates population for an initial state in the upper eigenstate). The separation between the two sites is $R = 20$, their coupling is $J = 0.25$, the speed of sound is $c = 2$, $\alpha = 0.12$ and $k_c = 1$. We can definitely see a revival of population at a time consistent with the amount of time needed for a bosonic excitation to travel into the bath from one system's site to the other.}
    \label{fig:unannotated}
\end{figure}

The unannotated version of the bath dynamics displayed in Fig.~\ref{fig:Upper_R=20_c=2_T=0}, presenting several consecutive revivals of the upper eigenstate population for $R = 20$ and $c = 2$ is shown in Fig.~\ref{fig:unannotated}.
\end{document}